\documentclass[journal,onecolumn,12pt]{IEEEtran}
\usepackage{amssymb}

\usepackage{cite}

\ifCLASSINFOpdf
  \usepackage[pdftex]{graphicx}
\else
  \usepackage[dvips]{graphicx}
\fi
\usepackage[cmex10]{amsmath}
\usepackage{algorithmic}

\usepackage{array}

\begin{document}
%
\title{EVM and Achievable Data Rate Analysis of Clipped OFDM Signals in Visible
Light Communication}
%
%
%

\author{Zhenhua~Yu,
        Robert~J.~Baxley,
        and~G.~Tong~Zhou
\thanks{Zhenhua Yu and G. Tong Zhou are with the School of Electrical and Computer Engineering, Georgia Institute of Technology, Atlanta,
GA 30332, USA, e-mail: zhenhuayu@gatech.edu.}
\thanks{Robert J. Baxley is with Georgia Tech Research Institute, Atlanta, GA 30332, USA.}
\thanks{This journal paper was published in EURASIP Journal on Wireless Communications and Networking 2012, 2012: 321. DOI: 10.1186/1687-1499-2012-321, October 2012 (url: http://jwcn.eurasipjournals.com/content/2012/1/321/abstract) \cite{zyu2012evm}.}}

%
%

%

\newtheorem{theorem}{Theorem}
\newtheorem{remark}[theorem]{Remark}



\maketitle

\begin{abstract}
Orthogonal frequency division multiplexing (OFDM) has been
considered for visible light communication (VLC); thanks to its
ability to boost data rates as well as its robustness against
frequency-selective fading channels. A major disadvantage of OFDM is
the large dynamic range of its time-domain waveforms, making OFDM
vulnerable to nonlinearity of light emitting diodes. DC-biased
optical OFDM (DCO-OFDM) and asymmetrically clipped optical OFDM
(ACO-OFDM) are two popular OFDM techniques developed for the VLC. In
this article, we will analyze the performance of the DCO-OFDM and
ACO-OFDM signals in terms of error vector magnitude (EVM),
signal-to-distortion ratio (SDR), and achievable data rates under
both average optical power and dynamic optical power constraints.
EVM is a commonly used metric to characterize distortions. We will
describe an approach to numerically calculate the EVM for DCO-OFDM
and ACO-OFDM. We will derive the optimum biasing ratio in the sense
of minimizing EVM for DCO-OFDM. In addition, we will formulate the
EVM minimization problem as a convex linear optimization problem and
obtain an EVM lower bound against which to compare the DCO-OFDM and
ACO-OFDM techniques. We will prove that the ACO-OFDM can achieve the
lower bound. Average optical power and dynamic optical power are two
main constraints in VLC. We will derive the achievable data rates
under these two constraints for both additive white Gaussian noise
channel and frequency-selective channel. We will compare the
performance of DCO-OFDM and ACO-OFDM under different power
constraint scenarios.
\end{abstract}


%
\IEEEpeerreviewmaketitle

\section{Introduction}
With rapidly growing wireless data demand and the saturation of
radio frequency (RF) spectrum, visible light communication (VLC)
\cite{komine2003integrated,komine2004fundamental,o2008visible,elgala2011indoor}
has become a promising candidate to complement conventional RF
communication, especially for indoor and medium range data
transmission. VLC uses white light emitting diodes (LEDs) which
already provide illumination and are quickly becoming the dominant
lighting source to transmit data. At the receiving end, a photo
diode (PD) or an image sensor is used as light detector. VLC has
many advantages including low-cost front-ends, energy-efficient
transmission, huge (THz) bandwidth, no electromagnetic interference,
no eye safety constraints like infrared, etc. \cite{elgala2007ofdm}.
In VLC, simple and low-cost intensity modulation and direct
detection (IM/DD) techniques are employed, which means that only the
signal intensity is modulated and there is no phase information. At
the transmitter, the white LED converts the amplitude of the
electrical signal to the intensity of the optical signal, while at
the receiver, the PD or image sensor generates the electrical signal
proportional to the intensity of the received optical signal. The
IM/DD requires that the electric signal must be real-valued and
unipolar (positive-valued).

Recently, orthogonal frequency division multiplexing (OFDM) has been
considered for VLC; thanks to its ability to boost data rates and
efficiently combat inter-symbol interference
\cite{hranilovic2005design,gonzalez2006adaptive,armstrong2006power,elgala2007ofdm,armstrong2009ofdm,
fernando2011flip}. To ensure that the OFDM time-domina signal is
real-valued, Hermitian symmetry condition must be satisfied in the
frequency-domain. Three methods have been discussed in the
literature for creating real-valued unipolar OFDM signal for VLC.

\begin{enumerate}
\item[(1)] DC-biased optical OFDM (DCO-OFDM)---adding a DC bias to the
original signal
\cite{carruthers1996multiple,hranilovic2005design,gonzalez2006adaptive};

\item[(2)] Asymmetrically clipped optical OFDM (ACO-OFDM)---only mapping
the data to the odd subcarriers and clipping the negative parts
without information loss \cite{armstrong2006power};

\item[(3)] Flip-OFDM---transmitting positive and negative parts in two
consecutive unipolar symbols \cite{fernando2011flip}.
\end{enumerate}
One disadvantage of OFDM is its high peak-to-average-power ratio
(PAPR) due to the summation over a large number of terms
\cite{tellado2000multicarrier}. The high PAPR or dynamic range of
OFDM makes it very sensitive to nonlinear distortions. In VLC, the
LED is the main source of nonlinearity. The nonlinear
characteristics of LED can be compensated by digital pre-distortion
(DPD) \cite{elgala2009non}, but the dynamic range of any physical
device is still limited. The input signal outside this range will be
clipped. A number of papers \cite{banelli2000theoretical,
ochiai2002performance,ochiai2003performance,peng2006capacity} have
studied the clipping effects on the RF OFDM signals. However,
clipping in the VLC system has two important differences: (i) the RF
baseband signal is complex-valued whereas time-domain signals in the
VLC system are real-valued; (ii) the main power limitation for VLC
is average optical power and dynamic optical power, rather than
average electrical power and peak power as in RF communication.
Therefore, most of the theory and analyses developed for RF OFDM are
not directly applicable to optical OFDM. A number of papers
\cite{armstrong2008comparison, elgala2009study,
elgala2009non,mesleh2011performance} have analyzed the LED
nonlinearity on DCO-OFDM and ACO-OFDM and compared their bit error
rate, power efficiency, bandwidth efficiency, etc.

In this article, we will investigate the performance of DCO-OFDM and
ACO-OFDM signals in terms of error vector magnitude (EVM),
signal-to-distortion ratio (SDR), and achievable data rates. EVM is
a frequently used performance metric in modern communication
standards. In \cite{elgala2009study, elgala2009non}, the EVM is
measured by simulations for varying power back-off and biasing
levels. In this article, we will describe an approach to numerically
calculate the EVM for DCO-OFDM and ACO-OFDM, and derive the optimum
biasing ratio for DCO-OFDM. We will formulate the EVM minimization
problem as a convex linear optimization problem and obtain an EVM
lower bound. In contrast to \cite{li2007channel} which investigated
the achievable data rates for ACO-OFDM with only average optical
power limitation, we will derive the achievable data rates subject
to both the average optical power and dynamic optical power
constraints. We will first derive the SDR for a given data-bearing
subcarrier based on the Bussgang's theory. Upon the SDR analysis, we
will derive the achievable data rates for additive white Gaussian
noise (AWGN) channel and frequency-selective channel. Finally, we
will compare the performance of two optical OFDM techniques.

\section{System model}

The system model discussed in this study is depicted in
Figure 1. In an OFDM system, a discrete time-domain
signal $\mathbf{x} = [x[0], x[1], \ldots, x[N-1]]$ is generated by
applying the inverse DFT (IDFT) operation to a frequency-domain
signal $\mathbf{X} = [X_0, X_1, \ldots, X_{N-1}]$ as
\begin{equation}
x[n] = \text{IDFT}(X_k) =
\frac{1}{\sqrt{N}}\sum_{k=0}^{N-1}X_k\exp(j2\pi kn/N), \quad 0\leq
n\leq N-1,
\end{equation}
where $j = \sqrt{-1}$ and $N$ are the size of IDFT, assumed to be an
even number in this article. In a VLC system using LED, the IM/DD
schemes require that the electric signal be real-valued and unipolar
(positive-valued). According to the property of IDFT, a real-valued
time-domain signal $x[n]$ corresponds to a frequency-domain signal
$X_k$ that is Hermitian symmetric, i.e.,
\begin{eqnarray}
\label{eq_herm}
X_k &=& X_{N-k}^*, \quad 1\leq k \leq N-1,\\\notag
X_k &\in& \mathbb{R}, \quad k = 0, N/2,
\end{eqnarray}
where $*$ denotes complex conjugate.

\begin{figure}[!t]
\begin{center}
\includegraphics[width=6.0in]{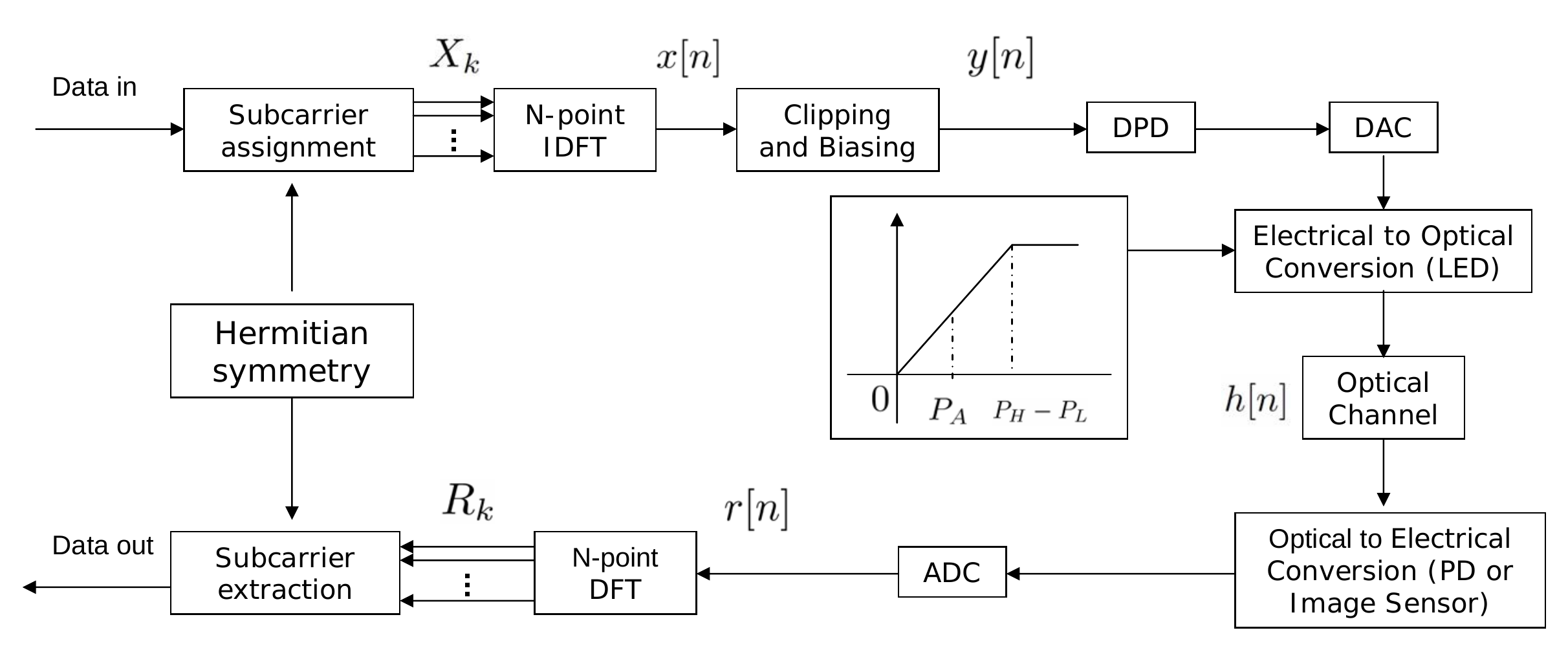}
\caption{OFDM system model in visible light communication}
\label{figmodel}
\end{center}
\end{figure}

In Figure 1, $y[n]$ is obtained from $x[n]$ after both
a clipping and a biasing operation are implemented. The resulting
signal, $y[n]$, is non-negative (i.e., $y[n] \geq 0$) and has a
limited dynamic range. In a VLC system, the light emitted is used
for illumination and communication simultaneously. The intensity of
light emitted by the LED is proportional to $y[n]$, while the
electrical power is proportional to $y^2[n]$. In the optical
communication literature, the average optical power of the LED input
signal $y[n]$ is defined as
\begin{equation}
\label{eqoydf}
O_{y} \triangleq \mathcal{E}\{y[n]\},
\end{equation}
where $\mathcal{E}[\cdot]$ denotes statistical expectation. Usually,
the VLC system operates under some average optical power constraint
$P_A$, i.e.,
\begin{equation}
O_{y} \leq P_A.
\end{equation}
This constraint is in place for two reasons: (i) the system power
consumption needs to be kept under a certain limit, (ii) the system
should still be able to communicate even under dim illumination
conditions. The VLC system is further limited by the dynamic range
of the LED. In this article, we assume that the DPD has perfectly
linearized the LED between the interval $[P_L, P_H]$, where $P_L$ is
the turn-on voltage (TOV) for the LED. If the TOV is provided by an
analog module at the LED and we assume that LED is already turned
on, the linear range for the input signal is $[0, P_H-P_L]$. We
define the dynamic optical power of $y[n]$ as
\begin{equation}
\label{eqgydf}
G_{y} \triangleq \max\left(y[n]\right)-\min\left(y[n]\right).
\end{equation}
$G_{y}$ should be constrained by $P_H-P_L$ as
\begin{equation}
G_{y} \leq P_H-P_L.
\end{equation}
Moreover, $y[n]$ must be non-negative, i.e., $y[n] \geq 0$.

According to the Central Limit Theorem, $x[n]$ is approximately
Gaussian distributed with zero mean and variance $\sigma^2$ with
probability density function (pdf):
\begin{equation}
p\left(x[n]=z\right)=
\frac{1}{\sigma}\phi\left(\frac{z}{\sigma}\right),
\end{equation}
where $\phi(x)= \frac{1}{\sqrt{2\pi}}e^{-\frac{1}{2}x^2}$ is the pdf of the standard Gaussian distribution. As a result, the time-domain OFDM signal $x[n]$ tends to occupy a large dynamic range and is bipolar.
In order to fit into the dynamic range of the LED, clipping is often
necessary, i.e.,
\begin{eqnarray}
\label{eq_clip}
\bar{x}[n]=\left\{\begin{array}{ccc}c_u,\,\, &&{x[n] > c_u }\\
x[n],\,\, &&{c_l  \leq x[n] \leq c_u }\\
c_l  ,\,\,&&  {x[n] < c_l}\end{array} \right.
\end{eqnarray}
where $c_u$ denotes the upper clipping level, and $c_l$ denotes the
lower clipping level. In order for the LED input $y[n]$ to be
non-negative, we may need to add a DC bias $B$ to the clipped signal
$\bar{x}[n]$ to obtain
\begin{eqnarray}
y[n] = \bar{x}[n] + B, \quad 0\leq n\leq N-1.
\end{eqnarray}
For $y[n] \geq 0$, we need $B=-c_l$.

To facilitate the analysis, we define the clipping ratio $\gamma$
and the biasing ratio $\varsigma$ as
\begin{equation}
\gamma \triangleq \frac{(c_u-c_l)/2}{\sigma}
\end{equation}
\begin{equation}
\varsigma \triangleq \frac{B}{c_u-c_l} = \frac{-c_l}{c_u-c_l}
\end{equation}
Thus, the upper and lower clipping levels can be written as
\begin{equation}
\label{eq_upper}
c_u = 2\sigma\gamma(1-\varsigma),
\end{equation}
\begin{equation}
\label{eq_lower}
c_l = -2\sigma\gamma\varsigma.
\end{equation}
The ratios $\gamma$ and $\varsigma$ can be adjusted independently
causing $c_u$ and $c_l$ to vary.

Clipping in the time-domain gives rise to distortions on all
subcarriers in the frequency domain. On the other hand, DC-bias only
affects the DC component in the frequency-domain. The clipped and
DC-biased signal $y[n]$ is then converted into analog signal and
subsequently modulate the intensity of the LED. At the receiver, the
photodiode, or the image sensor, converts the received optical
signal to electrical signal and transforms it to digital form. The
received sample can be expressed as
\begin{equation}
\label{eq_chcov}
r[n] =  (\bar{x}[n] + B) \otimes h[n] + w[n],
\end{equation}
where $h[n]$ is the impulse response of the wireless optical
channel, $w[n]$ is AWGN, and $\otimes$ denotes convolution. By
taking the DFT of Equation (\ref{eq_chcov}), we can obtain the
received data on the $k$th subcarrier as
\begin{equation}
\label{eq_chmul}
R_k =  \bar{X}_k H_k + W_k, \quad k \neq 0,
\end{equation}
where $H_k$ is the channel frequency response on the $k$th
subcarrier.

Based on the subcarrier arrangement, DC-biasing, or transmission
scheme, several optical OFDM techniques have been proposed in the
literature. In this article, we will focus on the performance
analysis of two widely studied optical OFDM techniques, namely,
DCO-OFDM and ACO-OFDM. In the following, we shall use superscripts
$^{(D)}$ and $^{(A)}$ to indicate DCO-OFDM and ACO-OFDM,
respectively.

In DCO-OFDM, subcarriers of the frequency-domain signal
$\mathbf{X}^{(D)}$ are arranged as
\begin{eqnarray}
\label{eq_dcoarrange} \mathbf{X}^{(D)} = [0 \quad X_1^{(D)} \quad
X_2^{(D)} \quad  \dots \quad X_{N/2-1}^{(D)} \quad  0 \quad
X_{N/2-1}^{*(D)} \quad \dots \quad X_2^{*(D)} \quad X_1^{*(D)}]
\end{eqnarray}
where the $0$th and $N/2$th subcarriers are null (do not carry
data). Equation (\ref{eq_dcoarrange}) reveals Hermitian symmetry
with respect to $k=N/2$. Let $\mathcal{K}_d$ denote the set of
data-carrying subcarriers with cardinality $|\mathcal{K}_d|$. The
set of data-carrying subcarriers for DCO-OFDM is
$\mathcal{K}_d^{(D)} = \{1, 2,\ldots,N/2-1, N/2+1,\ldots,N-2, N-1\}$
and $|\mathcal{K}_d^{(D)}| = N-2$. The time-domain signal
$x^{(D)}[n]$ can be obtained as
\begin{eqnarray}
x^{(D)}[n] =
\frac{2}{\sqrt{N}}\sum_{k=1}^{N/2-1}\left(\Re(X_{k}^{(D)})\cos(2\pi
kn/N)-\Im(X_{k}^{(D)})\sin(2\pi kn/N)\right),
\end{eqnarray}
which is real-valued. In DCO-OFDM, we first obtain a clipped signal
$\bar{x}^{(D)}[n]$ similar to the procedure in (\ref{eq_clip}), and
then add DC-bias $B = -c_l$ to obtain the LED input signal
\begin{eqnarray}
y^{(D)}[n] = \bar{x}^{(D)}[n] + B =
\left\{\begin{array}{ccl}c_u-c_l,\,\, &&{x^{(D)}[n] > c_u }\\
x^{(D)}[n] - c_l,\,\, &&{c_l  \leq x^{(D)}[n] \leq c_u }\\
0  ,\,\,&&  {x^{(D)}[n] < c_l}\end{array} \right.
\end{eqnarray}
In the frequency domain,
\begin{eqnarray}
Y_k^{(D)} = X_k^{(D)} + C_k, \quad \forall k \neq 0,
\end{eqnarray}
where $C_k$ is clipping noise on the $k$th subcarrier.

In ACO-OFDM, only odd subcarriers of the frequency-domain signal
$\mathbf{X^{(A)}}$ carry data
\begin{eqnarray}
\label{eq_acosub} \mathbf{X^{(A)}}  &=& [0 \quad X_1^{(A)} \quad 0
\quad X_3^{(A)} \quad \dots \quad 0 \quad X_{N/2-1}^{(A)} \quad  0
\quad X_{N/2-1}^{*(A)} \quad \dots \quad 0 \quad X_3^{*(A)} \quad 0
\quad  X_1^{*(A)}],
\end{eqnarray}
and $\mathbf{X^{(A)}}$ meets the Hermitian symmetry condition
(\ref{eq_herm}). The set of data-carrying subcarriers for ACO-OFDM
is $\mathcal{K}_d^{(A)} = \{1, 3,\ldots,N-1\}$ and
$|\mathcal{K}_d^{(A)}| = N/2$. Thus, the time-domain signal
$x^{(A)}[n]$ can be obtained as
\begin{eqnarray}
x^{(A)}[n]=
\frac{2}{\sqrt{N}}\sum_{q=0}^{N/4-1}\left(\Re(X_{2q+1}^{(A)})\cos\left(2\pi
(2q+1)n/N\right)-\Im(X_{2q+1}^{(A)})\sin\left(2\pi
(2q+1)n/N\right)\right),
\end{eqnarray}
which is real-valued. It follows easily that $x^{(A)}[n]$ satisfies
the following negative half symmetry condition:
\begin{equation}
\label{eqsymneg} x^{(A)}[n+N/2] = - x^{(A)}[n], \quad n = 0,
1,\ldots,N/2-1.
\end{equation}
Denote by $z[n]$ a generic discrete-time signal that satisfies
$z[n+N/2] = - z[n],\,n = 0, 1,\ldots,N/2-1$ and by $\bar{z}[n]$ its
clipped version where the negative values are removed, i.e.,
\begin{eqnarray}
\bar{z}[n]=\left\{\begin{array}{ccl}z[n],\,\, &&{z[n] \geq 0 },\\
0  ,\,\,&&  {\text{otherwise}}.\end{array} \right.
\end{eqnarray}
It was proved in \cite{wilson2009transmitter} that in the
frequency-domain,
\begin{equation}
\label{eq_zzkkhalf}
\bar{Z}_k = \frac{1}{2}Z_k,\quad \forall k \,\,{\text{odd}}.
\end{equation}
In ACO-OFDM, we obtain the LED input signal $y^{(A)}[n]$ via
\begin{eqnarray}
\label{eq_yclip}
y^{(A)}[n] = \left\{\begin{array}{ccl}c_u,\,\, &&{x^{(A)}[n] > c_u }\\
x^{(A)}[n],\,\, &&{0  \leq x^{(A)}[n] \leq c_u }\\
0  ,\,\,&&  {x^{(A)}[n] < 0}\end{array} \right.
\end{eqnarray}
Equation (\ref{eq_yclip}) can be regarded as a 2-step clipping
process, whereby we first remove those negative values in
$x^{(A)}[n]$, and then replace those $x^{(A)}[n]$ values that exceed
$c_u$ by $c_u$. Since $x^{(A)}[n]$ satisfies (\ref{eqsymneg}), we
infer based on (\ref{eq_zzkkhalf}) that
\begin{eqnarray}
\label{eq_acoyk}
Y_k^{(A)} = \frac{1}{2}X_k^{(A)} + C_k, \quad \forall k,
\end{eqnarray}
where $C_k$ is clipping noise on the $k$th subcarrier in the
frequency-domain. For ACO-OFDM, no DC-biasing is necessary and thus
the biasing ratio $\varsigma = 0$.

As an example, suppose that we need to transmit a sequence of eight
quadrature phase-shift keying (QPSK) symbols. Table 1
shows the subcarrier arrangement for DCO-OFDM, whereas
Table 2 shows the subcarrier arrangement for ACO-OFDM. The
time-main signals $x^{(D)}[n]$ and $x^{(A)}[n]$ and the
corresponding LED input signals $y^{(D)}[n]$ and $y^{(A)}[n]$ are
shown in{\break} Figure 2. We see that in $x^{(A)}[n]$, the
last 16 values are a repetition of the first 16 values but with the
opposite sign. It takes ACO-OFDM more bandwidth than DCO-OFDM to
transmit the same message, although ACO-OFDM is less demanding in
terms of dynamic range requirement of the LED and power consumption.
\begin{figure}
\begin{center}
\includegraphics[width=5.0in]{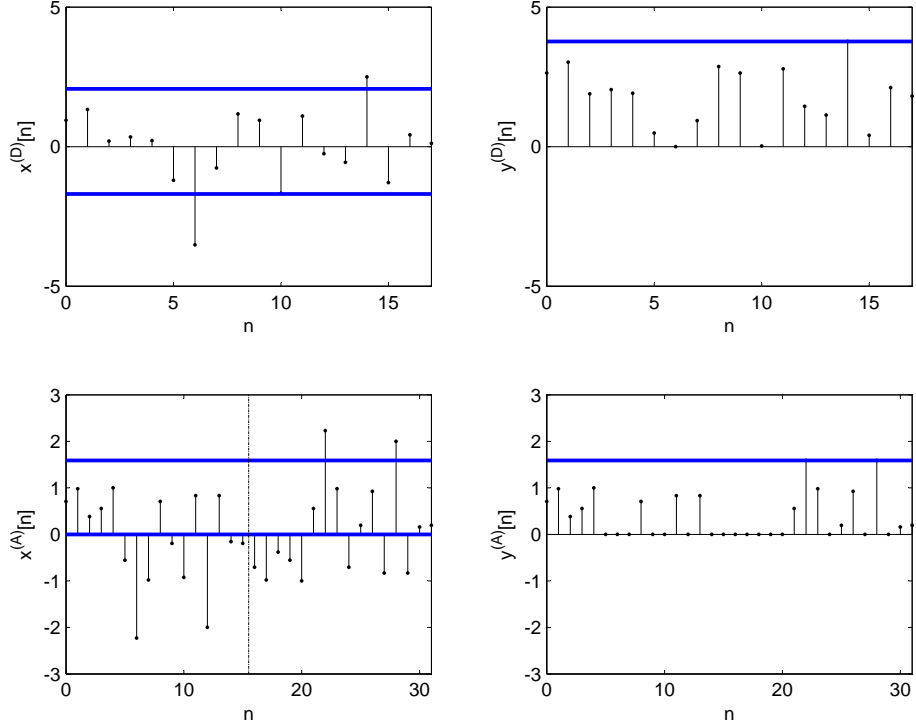}
\caption{An example of $x^{(D)}[n]$, $y^{(D)}[n]$, $x^{(A)}[n]$ and $y^{(A)}[n]$ to convey a sequence  8 QPSK symbols. For DCO-OFDM, $\gamma = 1.41 = 3$ dB, $\varsigma = 0.45$, $c_l = -1.70$, $c_u =  2.07$, $B = 1.70$; For ACO-OFDM, $\gamma = 0.79 = -2$ dB, $\varsigma = 0$, $c_l = 0$, $c_u = 1.59$, $B = 0$}
\label{fig_datime}
\end{center}
\end{figure}

\begin{table}[!h]
\subsection*{Table 1 DCO-OFDM subcarrier arrangement for transmitting eight
QPSK symbols---an example}\label{tab1}
\begin{tabular*}\textwidth{@{\extracolsep\fill}llllllllll@{\extracolsep\fill}}
\hline
$k$ &0& 1 & 2 & 3 & 4 & 5 & 6 & 7 & 8 \\
\hline
$X_k^{(D)}$ & 0 & $1+j$ & $1-j$ & $-1-j$ & $1+j$ & $1-j$ & $-1+j$ & $-1-j$ & $1-j$ \\
\hline
$k$ & 9 & 10 & 11 & 12 & 13 & 14 & 15 & 16 & 17 \\
\hline
$X_k^{(D)}$ & 0 & $1+j$ & $-1+j$ & $-1-j$ & $1+j$ & $1-j$ & $-1+j$ & $1+j$ & $1-j$\\
\hline
\end{tabular*}
\end{table}

\vspace*{12pt}
\begin{table}[!h]
\subsection*{Table 2 ACO-OFDM subcarrier arrangement for transmitting eight
QPSK symbols---an example}
\label{tab2}
\begin{tabular*}\textwidth{@{\extracolsep\fill}llllllllll@{\extracolsep\fill}}
\hline
$k$&0& 1 & 2& 3 &4& 5 &6& 7\\
\hline
$X_k^{(A)}$ & 0 & $1+j$ & 0 & $1-j$ & 0 & $-1-j$ & 0 & $1+j$  \\
$k$ &8 & 9 & 10 & 11 & 12 & 13 & 14 & 15\\
$X_k^{(A)}$ & 0 & $1-j$ & 0 & $-1+j$ & 0 & $-1-j$ & 0 & $1-j$ \\
$k$ &16& 17 & 18 & 19 & 20 & 21 & 22 & 23\\
$X_k^{(A)}$ & 0 & $1+j$ & 0 & $-1+j$ & 0 & $-1-j$ & 0 & $1+j$  \\
$k$ & 24 & 25 & 26 & 27 & 28 & 29 & 30 & 31 \\
$X_k^{(A)}$ & 0 & $1-j$ &0 & $-1+j$ & 0 & $1+j$ & 0 & $1-j$ \\
\hline
\end{tabular*}
\end{table}

\section{EVM analysis}
EVM is a figure-of-merit for distortions. Let $\mathbf{X}^{\dag}=
[X_0^{\dag}, X_1^{\dag}, \ldots, X_{N-1}^{\dag}]$ denote the
$N$-length DFT of the modified time-domain signal
$\mathbf{x}^{\dag}$. EVM can be defined as
\begin{equation}
\label{EVM} \xi(\mathbf{X}^{(r)}, \mathbf{X}^{\dag}) \triangleq
\sqrt{\frac{\mathcal{E}\left[\sum_{k \in \mathcal{K}_d}|X_k^{(r)} -
X_k^{\dag}|^2\right]}{\mathcal{E}\left[\sum_{k \in
\mathcal{K}_d}|X_k^{(r)}|^2\right]}},
\end{equation}
where $\mathbf{X}^{(r)}= [X_0^{(r)}, X_1^{(r)}, \ldots,
X_{N-1}^{(r)}]$ denotes the reference constellation. For DCO-OFDM,
$X_k^{(r)} = X_k^{(D)}$ for $k \in \mathcal{K}_d^{(D)}$. For
ACO-OFDM, $X_k^{(r)} = \frac{1}{2}X_k^{(A)}$ for $k \in
\mathcal{K}_d^{(A)}$.


\subsection{EVM calculation}
\label{sec_evm} In DCO-OFDM, clipping in the time-domain generates
distortions on all the subcarriers. We denote the clipping error
power by $\bar{P}^{(D)}_{\gamma,\varsigma}= \sum_{k \in
\mathcal{K}_d^{(D)}} \mathcal{E} [|X_k^{(D)} - \bar{X}_k^{(D)}|^2]$.
Since the sum distortion power on the 0th and $N/2$th subcarriers is
small relative to the total distortion power of $N$ subcarriers,
according to the Parseval's theorem, we can approximate
$\bar{P}^{(D)}_{\gamma,\varsigma}$ as
\begin{eqnarray}
\bar{P}^{(D)}_{\gamma,\varsigma} &=& \sum_{k \in
\mathcal{K}_d^{(D)}} \mathcal{E} [|X_k^{(D)} -
\bar{X}_k^{(D)}|^2]\\\notag &\approx& \sum_{n=0}^{N-1} \mathcal{E}
[|x^{(D)}[n] - \bar{x}^{(D)}[n]|^2]\\\notag &=&
N\left(\int_{c_u}^{\infty}(z-c_u)^2\frac{1}{\sigma}\phi\left(\frac{z}{\sigma}\right)dz
+
\int_{-\infty}^{c_l}(z-c_l)^2\frac{1}{\sigma}\phi\left(\frac{z}{\sigma}\right)dz\right)\\\notag
&=&
N\sigma^2\bigg(1+4\gamma^2(1-\varsigma)^2-2\gamma(1-\varsigma)\phi\big(2\gamma(1-\varsigma)\big)-2\gamma\varsigma\phi\big(2\gamma\varsigma\big)\\\notag
&&-\Phi\big(2\gamma(1-\varsigma)\big)-4\gamma^2(1-\varsigma)^2\Phi\big(2\gamma(1-\varsigma)\big)
+\Phi\big(-2\gamma\varsigma\big)+4\gamma^2\varsigma^2\Phi\big(-2\gamma\varsigma\big)\bigg),
\end{eqnarray}
where $\Phi(x)=\int_{-\infty}^x\phi(t)dt$.
Thus, we obtain the EVM for the DCO-OFDM scheme as
\begin{eqnarray}
\label{eqevmdco} \xi^{(D)}_{\gamma,\varsigma} &=&
\sqrt{\frac{\bar{P}^{(D)}_{\gamma,\varsigma}}{N\sigma^2}}\\\notag
&=&\bigg(1+4\gamma^2(1-\varsigma)^2-2\gamma(1-\varsigma)\phi\big(2\gamma(1-\varsigma)\big)-2\gamma\varsigma\phi\big(2\gamma\varsigma\big)
\\\notag
&&-\Phi\big(2\gamma(1-\varsigma)\big)-4\gamma^2(1-\varsigma)^2\Phi\big(2\gamma(1-\varsigma)\big)
+\Phi\big(-2\gamma\varsigma\big)+4\gamma^2\varsigma^2\Phi\big(-2\gamma\varsigma\big)\bigg)^{1/2}.
\end{eqnarray}
To find the optimum biasing ratio $\varsigma^{\star}$, we take the
first-order partial derivative and the second-order partial
derivative of $\bar{P}^{(D)}_{\gamma,\varsigma}$ with respect to the
biasing ratio $\varsigma$
\begin{eqnarray}
\frac{\partial\bar{P}^{(D)}_{\gamma,\varsigma}}{\partial\varsigma}
&=&
N\sigma^2\bigg(4\gamma\phi\big(2\gamma(1-\varsigma)\big)-4\gamma\phi\big(2\gamma\varsigma\big)\\\notag
&&-8\gamma^2(1-\varsigma)\Phi\big(2\gamma(\varsigma-1)\big)+8\gamma^2\varsigma\Phi\big(-2\gamma\varsigma\big)\bigg),\\
\frac{\partial^2\bar{P}^{(D)}_{\gamma,\varsigma}}{\partial\varsigma^2} &=& N\sigma^2\bigg(8\gamma^2\Phi\big(2\gamma(\varsigma-1)\big)+8\gamma^2\Phi\big(-2\gamma\varsigma\big)\bigg).
\end{eqnarray}
We can see that if $\varsigma = 0.5$,
$\partial\bar{P}^{(D)}_{\gamma,\varsigma}/\partial\varsigma=0$. The
second-order partial derivative
$\partial^2\bar{P}^{(D)}_{\gamma,\varsigma}/\partial\varsigma^2 > 0$
for all $\varsigma$. Hence, if $\varsigma < 0.5$,
$\partial\bar{P}^{(D)}_{\gamma,\varsigma}/\partial\varsigma<0$. If
$\varsigma > 0.5$,
$\partial\bar{P}^{(D)}_{\gamma,\varsigma}/\partial\varsigma>0$.
Therefore, $\varsigma^{\star} = 0.5$ is the optimum biasing ratio
which minimizes $\bar{P}^{(D)}_{\gamma,\varsigma}$. By substituting
$\varsigma^{\star}$ into Equation (\ref{eqevmdco})
we obtain the EVM for the DCO-OFDM scheme at the optimum biasing
ratio as
\begin{eqnarray}
\xi^{(D)}_{\gamma,\varsigma^*} &=&
\sqrt{\frac{\bar{P}^{(D)}_{\gamma,\varsigma^*}}{N\sigma^2}}\\\notag
&=&\sqrt{2(1+\gamma^2)\Phi(-\gamma)-2\gamma\phi(\gamma)}.
\end{eqnarray}

\begin{remark}
(i) $\varsigma^{\star} = 0.5$ is the optimum biasing ratio for
DCO-OFDM, regardless of the clipping ratio. (ii) When $\varsigma =
0.5$, we infer that $c_u = -c_l$, i.e., when the $x^{(D)}[n]$
waveform is symmetrically clipped at the negative and positive
tails, the clipping error power is always less than that when the
two tails are asymmetrically clipped (i.e., when $c_u \neq c_l$ or
when $\varsigma \neq 0.5$).
\end{remark}

Denote by $e[n]$, $n=0,1,\ldots,N-1$ a generic discrete-time signal
with DFT $E_k$, $k=0,1,\ldots,N-1$. When $k$ is odd, $E_k$ can be
written as
\begin{eqnarray}
\label{eq_ptnew1} E_k &=&
\frac{1}{\sqrt{N}}\sum_{n=0}^{N-1}e[n]\exp\left(-j2\pi\frac{kn}{N}\right)\\\nonumber
&=& \frac{1}{\sqrt{N}}\sum_{n=0}^{N/2-1}e[n]\exp\left(-j2\pi
\frac{kn}{N}\right) +
\frac{1}{\sqrt{N}}\sum_{n=0}^{N/2-1}e[n+N/2]\exp\left(-j2\pi
\frac{kn}{N}-jk\pi\right)\\\nonumber &=&
\frac{1}{\sqrt{N}}\sum_{n=0}^{N/2-1}\left(e[n]-e[n+N/2]\right)\exp\left(-j2\pi
\frac{kn}{N}\right).
\end{eqnarray}
Let $k = 2q+1$, $q = 0,1,\ldots,N/2-1$, Equation (\ref{eq_ptnew1})
can be further written as
\begin{eqnarray}
\label{eq_ptnew2} E_{2q+1} &=&
\frac{1}{\sqrt{N}}\sum_{n=0}^{N/2-1}\left(e[n]-e[n+N/2]\right)\exp\left(-j2\pi\frac{qn}{N/2}-j2\pi\frac{n}{N}\right)\\\nonumber
&=&\frac{1}{\sqrt{N}}\sum_{n=0}^{N/2-1}\exp\left(-j2\pi\frac{n}{N}\right)\left(e[n]-e[n+N/2]\right)\exp\left(-j2\pi\frac{qn}{N/2}\right)
\end{eqnarray}
Therefore, $\{E_k=E_{2q+1}\}_{q=0}^{N/2-1}$ can be viewed as the DFT
coefficients of a new discrete-time sequence
$\{\exp\left(-j2\pi\frac{n}{N}\right)\left(e[n]-e[n+N/2]\right)\}_{n=0}^{N/2-1}$.
Applying the Parseval's theorem to
$\{\exp\left(-j2\pi\frac{n}{N}\right)\left(e[n]-e[n+N/2]\right)\}_{n=0}^{N/2-1}$,
we obtain,
\begin{eqnarray}
\label{eq_ptnew3} \sum_{k = \text{odd}}|E_k|^2 =
\sum_{q=0}^{N/2-1}|E_{2q+1}|^2 =
\frac{1}{2}\sum_{n=0}^{N/2-1}\left(e[n]-e[n+N/2]\right)^2.
\end{eqnarray}

In ACO-OFDM, we denote the clipping error power by
$\bar{P}_{\gamma}^{(A)} = \sum_{k \in
\mathcal{K}_d^{(A)}}\mathcal{E}\left[\left|X_k^{(A)}/2-
\bar{X}_k^{(A)}\right|^2\right]$. According to (\ref{eq_ptnew3}), we
can calculate $\bar{P}_{\gamma}^{(A)}$ as
\begin{eqnarray}
\bar{P}^{(A)}_{\gamma} &=& \sum_{k \in
\mathcal{K}_d^{(A)}}\mathcal{E}\left[\left|X_k^{(A)}/2-
\bar{X}_k^{(A)}\right|^2\right]\\\notag &=&
\frac{1}{2}\sum_{n=0}^{N/2-1} \mathcal{E}
\left[\left(\frac{x^{(A)}[n]}{2} -
\bar{x}^{(A)}[n]-\frac{x^{(A)}[n+N/2]}{2} +
\bar{x}^{(A)}[n+N/2]\right)^2\right]\\\notag &=&
\frac{N}{4}\mathcal{E} \left[\left(x^{(A)}[n] - \bar{x}^{(A)}[n] +
\bar{x}^{(A)}[n+N/2]\right)^2\right]\\\notag &=&
\frac{N}{4}\int_{-\infty}^{-c_u}(z+c_u)^2\frac{1}{\sigma}\phi\left(\frac{z}{\sigma}\right)dz
+\frac{N}{4}\int_{c_u}^{\infty}(z-c_u)^2\frac{1}{\sigma}\phi\left(\frac{z}{\sigma}\right)dz\\\notag
&=&
\frac{N}{2}\sigma^2\bigg(-2\gamma\phi\left(2\gamma\right)+\Phi\left(-2\gamma\right)+4\gamma^2\Phi\left(-2\gamma\right)\bigg).
\end{eqnarray}
Then we obtain the EVM for the ACO-OFDM scheme as
\begin{eqnarray}
\xi^{(A)}_{\gamma} &=& \sqrt{\frac{\bar{P}^{(A)}_{\gamma}}{\sum_{k
\in
\mathcal{K}_d^{(A)}}\mathcal{E}\left[\left|X_k^{(A)}/2\right|^2\right]}}=
\sqrt{\frac{4\bar{P}^{(A)}_{\gamma,0}}{N\sigma^2}}\\\notag
&=&\sqrt{-4\gamma
\phi\left(2\gamma\right)+2\Phi\left(-2\gamma\right)+8\gamma^2\Phi\left(-2\gamma\right)}.
\end{eqnarray}

\subsection{Lower bound on the EVM}
Let us consider the setting
\begin{equation}
\label{eq_lb1}
\hat{x}[n] = x[n] + c[n],\quad 0 \leq n \leq N-1,
\end{equation}
where $x[n]$ is the original signal, $c[n]$ is a distortion signal,
and the resulting $\hat{x}[n]$ is expected to have a limited dynamic
range
\begin{equation}
\label{eq_lb2}
\max(\hat{x}[n]) - \min(\hat{x}[n]) \leq 2\gamma\sigma.
\end{equation}
In (\ref{eq_lb1}), all quantities involved are real-valued.

Clipping can produce one such $\hat{x}[n]$ signal, but there are
other less straightforward algorithms that can generate other
$\hat{x}[n]$ waveforms that also satisfy (\ref{eq_lb2}).

In the frequency-domain,
\begin{equation}
\label{eq_lb3}
\hat{X}_k = X_k + C_k.
\end{equation}
Since $x[n]$, $c[n]$, and $\hat{x}[n]$ are all real-valued, $X_k$,
$C_k$, and $\hat{X}_k$ all should satisfy the Hermitian symmetry
condition (\ref{eq_herm}). Therefore, $c[n]$ has the form
\begin{equation}
c[n]= \frac{2}{\sqrt{N}}\sum_{k=1}^{N/2-1}\bigg(\Re(C_k)\cos(2\pi
kn/N)-\Im(C_k)\sin(2\pi kn/N)\bigg) + \frac{1}{\sqrt{N}}C_0 +
\frac{1}{\sqrt{N}}C_{N/2}\cos(\pi n)
\end{equation}

We are interested in knowing the lowest possible EVM,
\begin{equation}
\xi =\sqrt{\frac{\mathcal{E}\left[\sum_{k \in
\mathcal{K}_d}|C_k|^2\right]}{\mathcal{E}\left[\sum_{k \in
\mathcal{K}_d}|X_k|^2\right]}}
\end{equation}
among all such $\hat{x}[n]$ waveforms. Afterwards, we can compare
the EVM from the DCO-OFDM and ACO-OFDM methods to get a sense of how
far these algorithms are from being optimum (in the EVM sense).

We formulate the following linear optimization problem:
\begin{equation}
\label{eq:convex}
\begin{aligned}
& {\text{minimize}}
& & \sum_{k \in \mathcal{K}_d} |C_k|^2\\
& \text{subject to}
& & \max\big(\hat{x}[n]\big) - \min\big(\hat{x}[n]\big)\leq 2\gamma\sigma\\
&
& &\hat{x}[n] = x[n] + \frac{2}{\sqrt{N}}\sum_{k=1}^{N/2-1}\bigg(\Re(C_k)\cos(2\pi kn/N)-\Im(C_k)\sin(2\pi kn/N)\bigg)\\
&
& & \quad \quad \quad \quad \quad +\frac{1}{\sqrt{N}}C_0+\frac{1}{\sqrt{N}}C_{N/2}\cos(\pi n) , \quad 0 \leq n  \leq N-1\\
&
& & C_0, C_{N/2} \in \mathbb{R}
\end{aligned}
\end{equation}

When the distortion of each OFDM symbol is minimized by the above
convex optimization approach, the corresponding EVM of $\hat{x}[n]$
(which is proportional to $\sqrt{\mathcal{E}\left[\sum_{k \in
\mathcal{K}_d} |C_k|^2\right]}$) serves as the lower bound for the
given dynamic range $2\gamma\sigma$.

\subsection{Optimality for ACO-OFDM}
\label{sec_opt} In this section, we will prove that the ACO-OFDM
scheme achieves the minimum EVM and thus is optimal in the EVM
sense.

For ACO-OFDM, let us write
\begin{equation}
\hat{x}[n] = x^{(A)}[n] + c[n],
\end{equation}
where $c[n]$ is the clipping noise to ensure that $\hat{x}[n]$ has a
limited dynamic range as described in (\ref{eq_lb2}). In the
frequency-domain, we have
\begin{equation}
\hat{X}_k = X_k^{(A)} + C_k
\end{equation}
where the $X_k^{(A)}$ subcarriers are laid out as in
(\ref{eq_acosub}). According to (\ref{eq_ptnew3}), when $k$ is odd,
the objective function in (\ref{eq:convex}) can be written as
\begin{eqnarray}
\label{eq_acoopt01} \sum_{k = \text{odd}} |C_k|^2 &=&
\frac{1}{2}\sum_{n=0}^{N/2-1}(c[n]-c[n+N/2])^2.
\end{eqnarray}
The dynamic range constraints in problem (\ref{eq:convex}) can be
viewed as two constraints put together.
\begin{equation}
\label{eq_acoopt11} \max\big(x^{(A)}[n]+ c[n]\big) -
\min\big(x^{(A)}[n] + c[n]\big) \leq 2\gamma\sigma,\quad 0 \leq n
\leq N/2-1,
\end{equation}
\begin{equation}
\label{eq_acoopt12} \max\big(x^{(A)}[n] + c[n]\big)
-\min\big(x^{(A)}[n] + c[n]\big) \leq 2\gamma\sigma, \quad N/2 \leq
n \leq N-1.
\end{equation}
Since $x^{(A)}[n] = -x^{(A)}[n-N/2]$ when $N/2 \leq n \leq N-1$,
Equation (\ref{eq_acoopt12}) can be further written as
\begin{equation}
\label{eq_acoopt13} \max\big(x^{(A)}[n]-c[n+N/2]\big) -
\min\big(x^{(A)}[n] - c[n+N/2]\big) \leq 2\gamma\sigma, \quad 0 \leq
n \leq N/2-1.
\end{equation}
From Equations (\ref{eq_acoopt01}), (\ref{eq_acoopt11}), and
(\ref{eq_acoopt13}),  the problem (\ref{eq:convex}) can be recast as
\begin{equation}
\begin{aligned}
& {\text{minimize}}
& & \sum_{n=0}^{N/2-1}c^2[n]\\
& \text{subject to} & & \max\big(x^{(A)}[n] + c[n]\big) -
\min\big(x^{(A)}[n] + c[n]\big) \leq 2\gamma\sigma, \quad 0 \leq n
\leq N/2-1\\
\end{aligned}
\end{equation}
which is equivalent to
\begin{equation}
\label{eq_finalopt}
\begin{aligned}
& {\text{minimize}}
& & \sum_{n=0}^{N/2-1}c^2[n]\\
& \text{subject to}
& & x^{(A)}[n] + c[n]\leq 2\gamma\sigma, \quad 0 \leq n  \leq N/2-1\\
&
& & x^{(A)}[n] + c[n]\geq 0, \quad 0 \leq n  \leq N/2-1\\
\end{aligned}
\end{equation}
In Appendix, we prove that the solution $c^{\star}[n]$ to
(\ref{eq_finalopt}) yields
\begin{eqnarray}
\bar{x}^{(A)}[n] = x^{(A)}[n]+ c^{\star}[n] = \left\{\begin{array}{ccl}2\gamma\sigma,\,\, &&{x^{(A)}[n] > 2\gamma\sigma }\\
x^{(A)}[n],\,\, &&{0  \leq x^{(A)}[n] \leq 2\gamma\sigma }\\
0  ,\,\,&&  {x^{(A)}[n] < 0}\end{array} \right.
\end{eqnarray}
and thus the ACO-OFDM scheme is optimum in the EVM sense.

\section{SDR analysis}
Based on the Bussgang's theorem \cite{bussgang1952crosscorrelation},
any nonlinear function of $x[n]$ can be decomposed into a scaled
version of $x[n]$ plus a distortion term $d[n]$ that is uncorrelated
with $x[n]$. For example, we can write
\begin{equation}
\label{bueq} \bar{x}[n] = \alpha\cdot x[n] + d[n],\quad
n=0,\ldots,N-1.
\end{equation}
Let $R_{xx}[m] = \mathcal{E}\{x[n]x[n+m]\}$ denote the
auto-correlation function of $x[n]$, and let $R_{xy}[m] =
\mathcal{E}\{x[n]y[n+m]\}$ denote the cross-correlation function
between $x[n]$ and $y[n]$ at lag $m$. For any given $m$, the
correlation functions satisfy
\begin{equation}
R_{xd}[m] = 0,
\end{equation}
\begin{equation}
R_{\bar{x}x}[m] = \alpha R_{xx}[m].
\end{equation}
Thus, the scaling factor $\alpha$ can be calculated as
\begin{eqnarray}
\label{eq_alpha}
\alpha &=& \frac{R_{\bar{x}x}[0]}{ R_{xx}[0]}\\\notag
&=& \frac{\mathcal{E}\{\bar{x}[n]x[n]\}}{\sigma^2}\\\notag
&=& \frac{1}{\sigma^2}\int_{-\infty}^{\infty}\bar{x}x\cdot p(x) dx
\end{eqnarray}
Let $f(\cdot)$ denote the function linking the original signal to
the clipped signal, it is shown in \cite{davenport1958introduction}
that the output auto-correlation function $R_{\bar{x}\bar{x}}[m]$ is
related to the input auto-correlation function $R_{xx}[m]$ via
\begin{eqnarray}
\label{ry} R
_{\bar{x}\bar{x}}[m]=\sum_{\ell=0}^{\infty}\frac{b_{\ell}^2}{\ell!}\left[\frac{R_{xx}[m]}{\sigma^2}\right]^{\ell},
\end{eqnarray}
where the coefficients
\begin{eqnarray}
\label{coeexp} b_{\ell}=
\frac{(-1)^{\ell}\sigma^{\ell-1}}{\sqrt{2\pi}}\int_{-\infty}^{\infty}f(x)\frac{d^{\ell}[\exp(-\frac{x^2}{2\sigma^2})]}{dx^{\ell}}dx.
\end{eqnarray}
The input auto-correlation function $R_{xx}[m]$ can be obtained from
taking IDFT of the input power spectrum density (PSD)
\begin{eqnarray}
R_{xx}[m]= \texttt{IDFT}\{P_{X,k}\}_m, \quad m=0,\ldots,N-1,
\end{eqnarray}
where $P_{X,k} = \mathcal{E}[|X_k|^2]$ is the expected value of the
power on the $k$th subcarrier before clipping. Then it is
straightforward to calculate the output PSD by taking the DFT of the
auto-correlation of the output signal:
\begin{eqnarray}
P_{\bar{X},k}=\texttt{DFT}\{R_{\bar{x}\bar{x}}[m]\}_k, \quad
k=0,\ldots,N-1.
\end{eqnarray}
Taking the DFT of Equation (\ref{bueq}), the data at the $k$th
subcarrier are expressed as
\begin{eqnarray}
\bar{X}_k &=& \texttt{DFT}\{\alpha\cdot x[n]\}_k + \texttt{DFT}\{d[n]\}_k\\\notag
&=& \alpha\cdot X_k + D_k, \quad k \in \mathcal{K}_d.
\end{eqnarray}
Here, we assume that $D_k$ is Gaussian distributed, which is the
common assumption when $N$ is large \cite{ochiai2002performance}.
The SDR at the $k$th subcarrier is given by
\begin{equation}
\texttt{SDR}_k = \frac{\mathcal{E}[|\alpha\cdot
X_k|^2]}{\mathcal{E}[|D_k|^2]} = \frac{\alpha^2P_{X,k}}{P_{D,k}} =
\frac{\alpha^2P_{X,k}}{P_{\bar{X},k}-\alpha^2P_{X,k}}, \quad k \in
\mathcal{K}_d,
\end{equation}
where $P_{D,k}=\mathcal{E}[|D_k|^2] = P_{\bar{X},k}-\alpha^2P_{X,k}$
is the average power of the distortion on the $k$th subcarrier.

According to Equation (\ref{eq_alpha}), we can obtain the scaling
factor $\alpha$ as a function of the clipping ratio $\gamma$ and the
biasing ratio $\varsigma$:
\begin{eqnarray}
\label{eq_llll} \alpha &=&
\frac{1}{\sigma^2}\int_{-\infty}^{\infty}\bar{x}x\cdot p(x)
dx\\\notag &=&
\frac{1}{\sigma^2}\int_{c_l}^{c_u}z^2\frac{1}{\sigma}\phi\left(\frac{z}{\sigma}\right)dz
+
\frac{1}{\sigma^2}\int_{-\infty}^{c_l}c_lz\frac{1}{\sigma}\phi\left(\frac{z}{\sigma}\right)dz
+
\frac{1}{\sigma^2}\int_{c_u}^{\infty}c_uz\frac{1}{\sigma}\phi\left(\frac{z}{\sigma}\right)dz\\\notag
&=& \Phi\left(2\gamma(1-\varsigma)\right) -
\Phi\left(-2\gamma\varsigma\right).
\end{eqnarray}
Note that in (\ref{eq_llll}) we have used Equations (\ref{eq_lower})
and (\ref{eq_upper}) for $c_l$ and $c_u$. According to Equation
(\ref{coeexp}), we can obtain the coefficient $b_{\ell}$ as a
function of the clipping ratio $\gamma$ and the biasing ratio
$\varsigma$:
{\small\begin{eqnarray}
b_{\ell}=\left\{\begin{array}{ccl}
\sigma\phi(2\gamma\varsigma)-\sigma\phi\left(2\gamma(1-\varsigma)\right)-2\sigma\gamma\varsigma\Phi\left(-2\gamma\varsigma\right)+2\sigma\gamma(1-\varsigma)
\Phi\left(-2\gamma(1-\varsigma)\right),\,\, && {\ell=0}\\
\\
\sigma\Phi\left(2\gamma(1-\varsigma)\right)-\sigma\Phi(-2\gamma\varsigma),\,\, &&{\ell = 1 }\\
\\
\frac{\sigma}{\sqrt{2\pi}}\exp{\left(-2\gamma^2\varsigma^2\right)} He_{(\ell-2)}\left(-2\gamma\varsigma\right)-\frac{\sigma}{\sqrt{2\pi}}\exp{\left(-2\gamma^2(1-\varsigma)^2\right)} He_{(\ell-2)}\left(2\gamma(1-\varsigma)\right),\,\, &&{\ell >1 }
\end{array} \right.
\end{eqnarray}}
where $He_{n}(t) =
(-1)^{\ell}\exp{\left(\frac{t^2}{2}\right)}\frac{d^{\ell}[\exp(-\frac{t^2}{2})]}{dt^{\ell}}$
is the probabilists' Hermite polynomials
\cite{abramowitz1964handbook}.

\section{Achievable data rate}

In VLC, average optical power and dynamic optical power are two main
constraints.
Recall from Equation (\ref{eqoydf}), we can obtain the average
optical power of $y[n]$ as
\begin{eqnarray}
\label{eqayd}
O_{y} &=& \mathcal{E}\{y[n]\}\\\notag
&=& \mathcal{E}\{\bar{x}[n]\}+ B\\\notag
&=& \int_{c_l}^{c_u}z\frac{1}{\sigma}\phi\left(\frac{z}{\sigma}\right)dz + c_u\int_{c_u}^{\infty}\frac{1}{\sigma}\phi\left(\frac{z}{\sigma}\right)dz+ c_l\int_{-\infty}^{c_l}\frac{1}{\sigma}\phi\left(\frac{z}{\sigma}\right)dz - c_l\\\notag
&=& \sigma\bigg(\phi(2\gamma\varsigma)-\phi\left(2\gamma(1-\varsigma)\right)-2\gamma\varsigma\Phi\left(-2\gamma\varsigma\right)+2\gamma(1-\varsigma)\Phi\left(-2\gamma(1-\varsigma)\right)+2\gamma\varsigma\bigg).
\end{eqnarray}
Let $\sigma^2_w = \mathcal{E}\{w^2[n]\}$ denote the power of AWGN
$w[n]$, we define the optical signal-to-noise ratio (OSNR) as
\begin{equation}
\texttt{OSNR} =  \frac{O_{y}}{\sigma_w}.
\end{equation}
Recall from Equation (\ref{eqgydf}), we can obtain the dynamic
optical power of $y[n]$ as
\begin{equation}
\label{eqgy} G_{y} = \max\left(y[n]\right)-\min\left(y[n]\right) =
c_u-c_l = 2\sigma\gamma.
\end{equation}
We define the dynamic signal-to-noise ratio (DSNR) as
\begin{equation}
\texttt{DSNR} =  \frac{G_{y}}{\sigma_w}.
\end{equation}
Let $\eta_{\rm OSNR} = P_A/\sigma_w$ denote the OSNR constraint and
$\eta_{\rm DSNR} = (P_H-P_L)/\sigma_w$ denote the DSNR constraint,
we have
\begin{equation}
\label{eq_less1} \frac{\sigma}{\sigma_w} \leq \frac{\eta_{\rm
OSNR}}{O_{y}/\sigma},
\end{equation}
\begin{equation}
\label{eq_less2} \frac{\sigma}{\sigma_w} \leq \frac{\eta_{\rm
DSNR}}{G_{y}/\sigma}.
\end{equation}
The maximum $\sigma/\sigma_w$ value can be obtained as
\begin{equation}
\label{eq_maxsw} \frac{\sigma}{\sigma_w}= \min\left(\frac{\eta_{\rm
OSNR}}{\phi(2\gamma\varsigma)-\phi\left(2\gamma(1-\varsigma)\right)-2\gamma\varsigma
\Phi\left(-2\gamma\varsigma\right)+2\gamma(1-\varsigma)\Phi\left(-2\gamma(1-\varsigma)
\right)+2\gamma\varsigma},\quad\frac{\eta_{\rm
DSNR}}{2\gamma}\right),
\end{equation}
by substituting (\ref{eqayd}) and (\ref{eqgy}) into the right-hand
side of (\ref{eq_less1}) and (\ref{eq_less2}), respectively. The
ratio $\eta_{\rm DSNR}/\eta_{\rm OSNR}= (P_H-P_L) / P_A$ is
determined by specific system requirements.

\subsection*{AWGN channel}
For AWGN channel, recall from Equation (\ref{eq_chmul}), the
received data on the $k$th subcarrier can be expressed as
\begin{equation}
R_k = \bar{X}_k + W_k = \alpha X_k + D_k + W_k, \quad k \in
\mathcal{K}_d.
\end{equation}
The signal-to-noise-and-distortion ratio (SNDR) for the $k$th
subcarrier is given by
\begin{eqnarray}
\label{eqsndr} \texttt{SNDR}_k &=&
\frac{\alpha^2\mathcal{E}\{|X_k|^2\}}{\mathcal{E}\{|D_k|^2\}+\mathcal{E}\{|W_k|^2\}}\\\notag
&=& \frac{\alpha^2P_{X,k}}{P_{D,K}+\sigma^2_w}\\\notag &=&
\frac{1}{\texttt{SDR}_k^{-1} +
\sigma_w^2\cdot\frac{1}{\alpha^2P_{X,k}}}.
\end{eqnarray}
In this article, we assume the power is equally distributed on all
data-carrying subcarriers,
\begin{eqnarray}
\label{eqpowerdis}
P_{X,k} = \frac{N\sigma^2}{|\mathcal{K}_d|},
\end{eqnarray}
then Equation (\ref{eqsndr}) is reduced to
\begin{eqnarray}
\label{eqsndras}
\texttt{SNDR}_k = \frac{1}{\texttt{SDR}_k^{-1} + \frac{\sigma_w^2}{\sigma^2}
\cdot\frac{|\mathcal{K}_d|}{N\alpha^2}}.
\end{eqnarray}
By substituting Equation (\ref{eq_maxsw}) into (\ref{eqsndras}), we
obtain the reciprocal of SNDR at the $k$th subcarrier:
{\small\begin{eqnarray}
&&\left(\texttt{SNDR}_k\right)^{-1} =
\left(\texttt{SDR}_k\right)^{-1} +
\frac{|\mathcal{K}_d|}{N\alpha^2}\\\nonumber
&&\cdot\max\left(\frac{\bigg(\phi(2\gamma\varsigma)-\phi\left(2\gamma(1-\varsigma)\right)-2\gamma\varsigma\Phi\left(-2\gamma\varsigma\right)+2\gamma(1-\varsigma)\Phi\left(-2\gamma(1-\varsigma)\right)+2\gamma\varsigma\bigg)^2}{\eta_{\rm
OSNR}^2},\quad \frac{4\gamma^2}{\eta_{\rm DSNR}^2}\right).
\end{eqnarray}}
Therefore, the achievable data rate, as a function of clipping ratio
$\gamma$, $\varsigma$, $\eta_{\rm OSNR}$, and $\eta_{\rm DSNR}$, is
given by
\begin{eqnarray}
\mathcal{R}\left(\gamma, \varsigma, \eta_{\rm OSNR}, \eta_{\rm
DSNR}\right) = \frac{1}{2N}\sum_{k\in \mathcal{K}_d}\log_2\left(1+
\texttt{SNDR}_k\right) \frac{\texttt{bits}}{\texttt{subcarrier}}.
\end{eqnarray}

\subsection*{Frequency-selective channel}
In the presence of frequency-selective channel, the received data on
the $k$th subcarrier obey the following in the frequency-domain:
\begin{equation}
\label{receivch} R_k =  H_k\bar{X}_k + W_k = H_k(\alpha X_k + D_k) +
W_k, \quad k \in \mathcal{K}_d.
\end{equation}
In this article, we consider the ceiling bounce channel model
\cite{kahn1997wireless} given by
\begin{equation}
h(t) = H(0)\frac{6a^6}{(t+a)^7}u(t),
\end{equation}
where $H(0)$ is the gain constant, $a = 12\sqrt{11/23}D$ and $u(t)$
is the unit step function. $D$ denotes the rms delay. From Equation
(\ref{receivch}), the SNDR is given by
\begin{eqnarray}
\label{eqsndrsl}
\texttt{SNDR}_k (H_k)&=& \frac{|H_k|^2\alpha^2\mathcal{E}\{|X_k|^2\}}{|H_k|^2\mathcal{E}\{|D_k|^2\}+\mathcal{E}\{|W_k|^2\}}\\
&=& \frac{1}{\texttt{SDR}_k^{-1} + \sigma_w^2\cdot\frac{1}{|H_k|^2\alpha^2P_{X,k}}}.
\end{eqnarray}
With the assumption of equal power distribution, we can obtain the
$1/\texttt{SNDR}$ as
{\small\begin{eqnarray}
\label{sndrchdco} &&\left(\texttt{SNDR}_k\right)^{-1} =
\left(\texttt{SDR}_k\right)^{-1} +
\frac{|\mathcal{K}_d|}{N|H_k|^2\alpha^2}\\\nonumber
&&\cdot\max\left(\frac{\bigg(\phi(2\gamma\varsigma)-\phi\left(2\gamma(1-\varsigma)\right)-2\gamma\varsigma\Phi\left(-2\gamma\varsigma\right)+2\gamma(1-\varsigma)\Phi\left(-2\gamma(1-\varsigma)\right)+2\gamma\varsigma\bigg)^2}{\eta_{\rm
OSNR}^2},\quad \frac{4\gamma^2}{\eta_{\rm DSNR}^2}\right).
\end{eqnarray}}
The achievable data rate, in the presence of frequency-selective
channel, is given by
\begin{eqnarray}
\label{eqdachrate} \mathcal{R}\left(\gamma, \varsigma, \eta_{\rm
OSNR}, \eta_{\rm DSNR}, \mathbf{H}\right) = \frac{1}{2N}\sum_{k\in
\mathcal{K}_d}\log_2\left(1+ \texttt{SNDR}_k(H_k)\right)
\frac{\texttt{bits}}{\texttt{subcarrier}}.
\end{eqnarray}

\section{Numerical results}
In this section, we show EVM simulation results and achievable data
rates of clipped optical OFDM signals under various average optical
power and dynamic optical power constraints.

\subsection{EVM simulation}
The EVM analyses for DCO-OFDM and ACO-OFDM are validated through
computer simulations. In the simulations, we chose the number of
subcarriers $N=512$, and QPSK modulation. One thousand OFDM symbols
were generated based on which we calculated the EVM. In order to
experimentally determine the optimum biasing ratio for DCO-OFDM, we
used biasing ratios ranging from 0.3 to 0.7 in step size of 0.02,
and clipping ratios ranging from 5 to 9\,dB in step size of 1\,dB.
Their simulated and theoretical EVM curves are plotted in
Figure 3. As expected, the minimum EVM was achieved
when the biasing ratio was 0.5, regardless of the clipping ratio.
This agrees with the analysis in ``EVM calculation'' section.
Next, we compared the EVM for DCO-OFDM with biasing ratio 0.5, EVM
for ACO-OFDM with biasing ratio 0, and their respective lower
bounds. To obtain the lower bounds, we used \texttt{CVX}, a package
for specifying and solving convex programs \cite{CVX}, to solve
Equation (\ref{eq:convex}).
The resulting EVM curves for DCO-OFDM are plotted in
Figure 4. The resulting EVM curves for ACO-OFDM are
plotted in Figure 5. We see that the EVM for ACO-OFDM
achieves its lower bound, thus corroborating the discussion in
``Optimality for ACO-OFDM'' section. For DCO-OFDM, the gap above the
lower bound increases with the clipping ratio (i.e., with increasing
dynamic range of the LED). This implies that there exists another
(more complicated) way of mapping $x^{D}[n]$ into a limited dynamic
range signal $\hat{x}[n]$ that can yield a lower EVM.

\begin{figure}[!t]
\begin{center}
\includegraphics[width=5.0in]{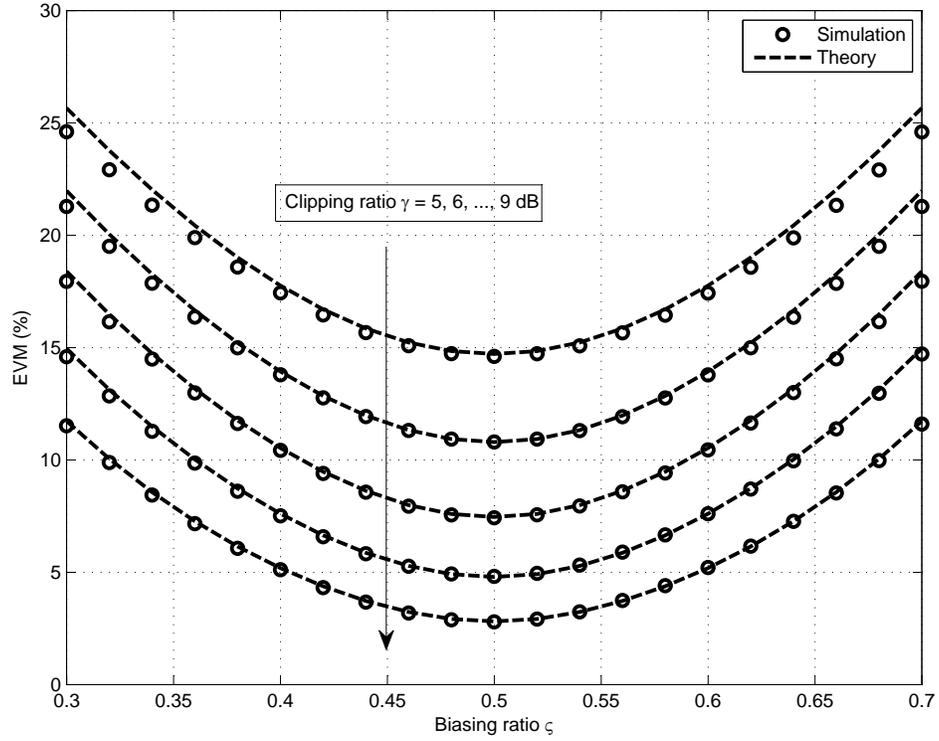}
\caption{EVM as a function of biasing ratio for DCO-OFDM with clipping ratio = 5, 6, ..., 9 dB}
\label{figdcoevm}
\end{center}
\end{figure}

\begin{figure}[!t]
\begin{center}
\includegraphics[width=5.0in]{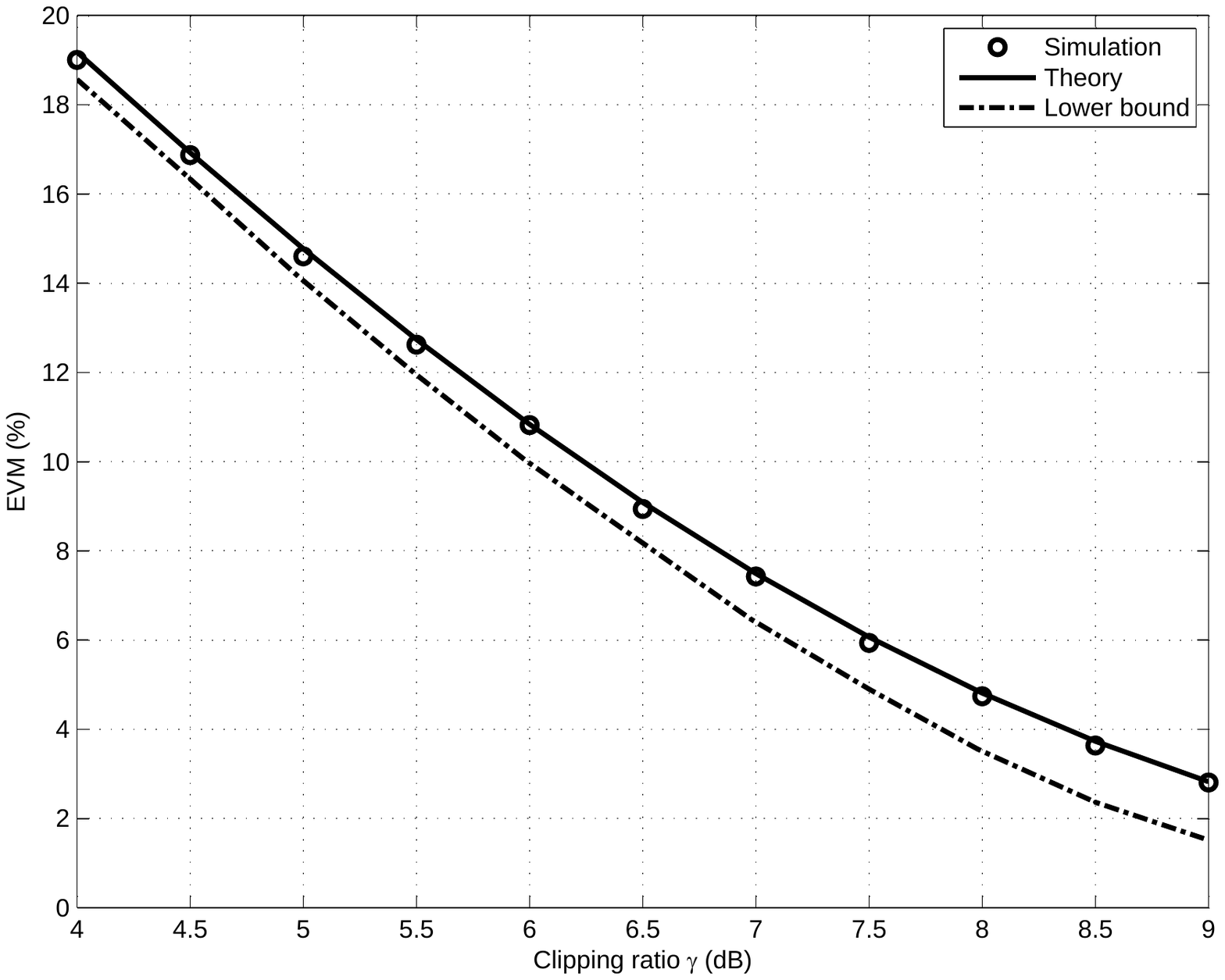}
\caption{EVM as a function of the clipping ratio $\gamma$ for DCO-OFDM along with the EVM lower bound for a given dynamic range limit $2\gamma\sigma$.}
\label{figevmdco}
\end{center}
\end{figure}

\begin{figure}[!t]
\begin{center}
\includegraphics[width=5.0in]{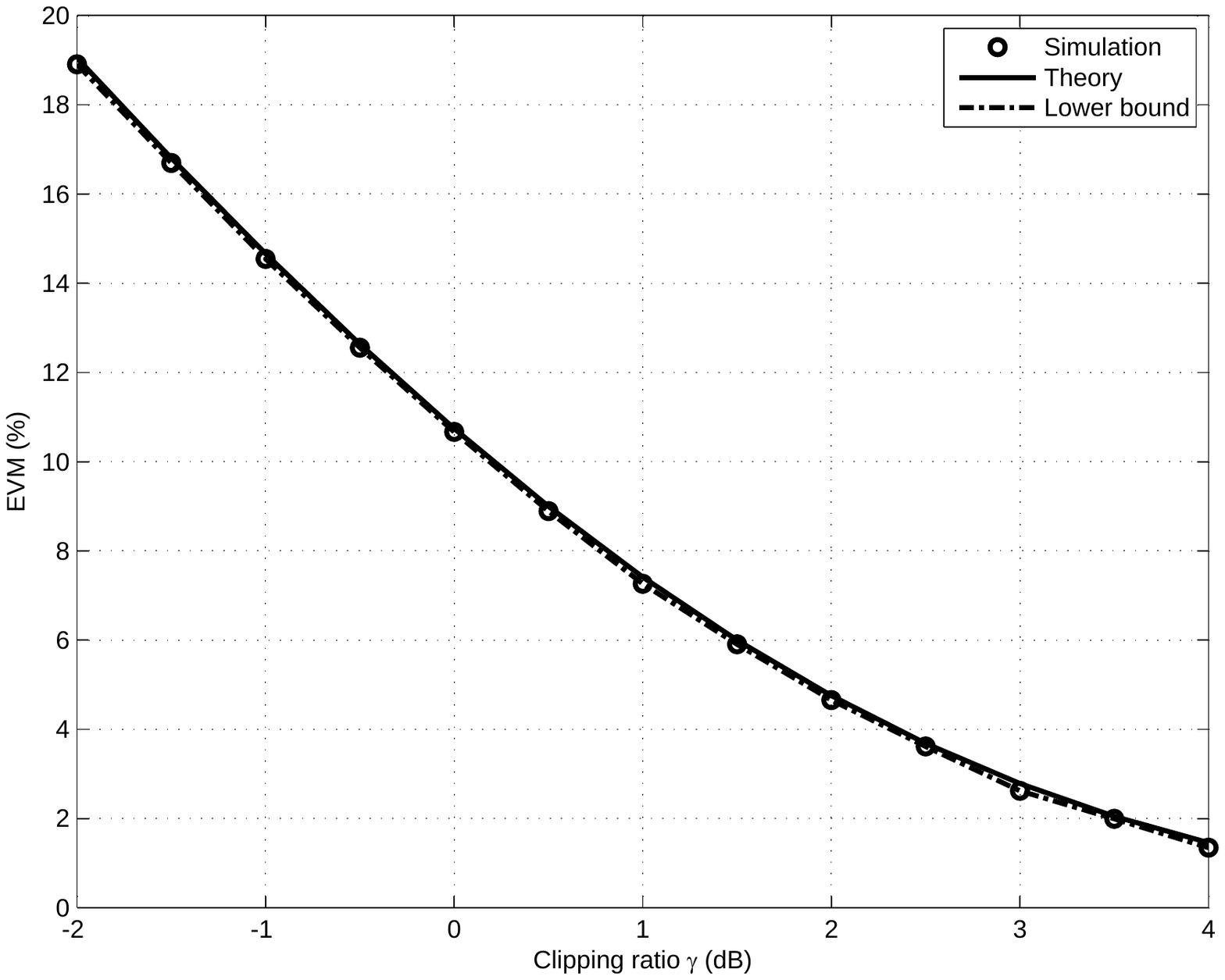}
\caption{EVM as a function of the clipping ratio $\gamma$ for ACO-OFDM along with the EVM lower bound for a given dynamic range limit $2\gamma\sigma$. }
\label{figevmaco}
\end{center}
\end{figure}


\subsection*{Achievable data rates performance}
We now show achievable data rates of clipped OFDM signals under
various average optical power and dynamic optical power constraints.
The number of subcarriers was $N=512$. For the frequency-selective
channel, we chose the rms delay spread $D = 10$\,ns and sampling
frequency 100\,MHz. The normalized frequency response for each
subcarrier is shown in Figure 6.
\begin{figure}[htc]
\begin{center}
\includegraphics[width=5.0in]{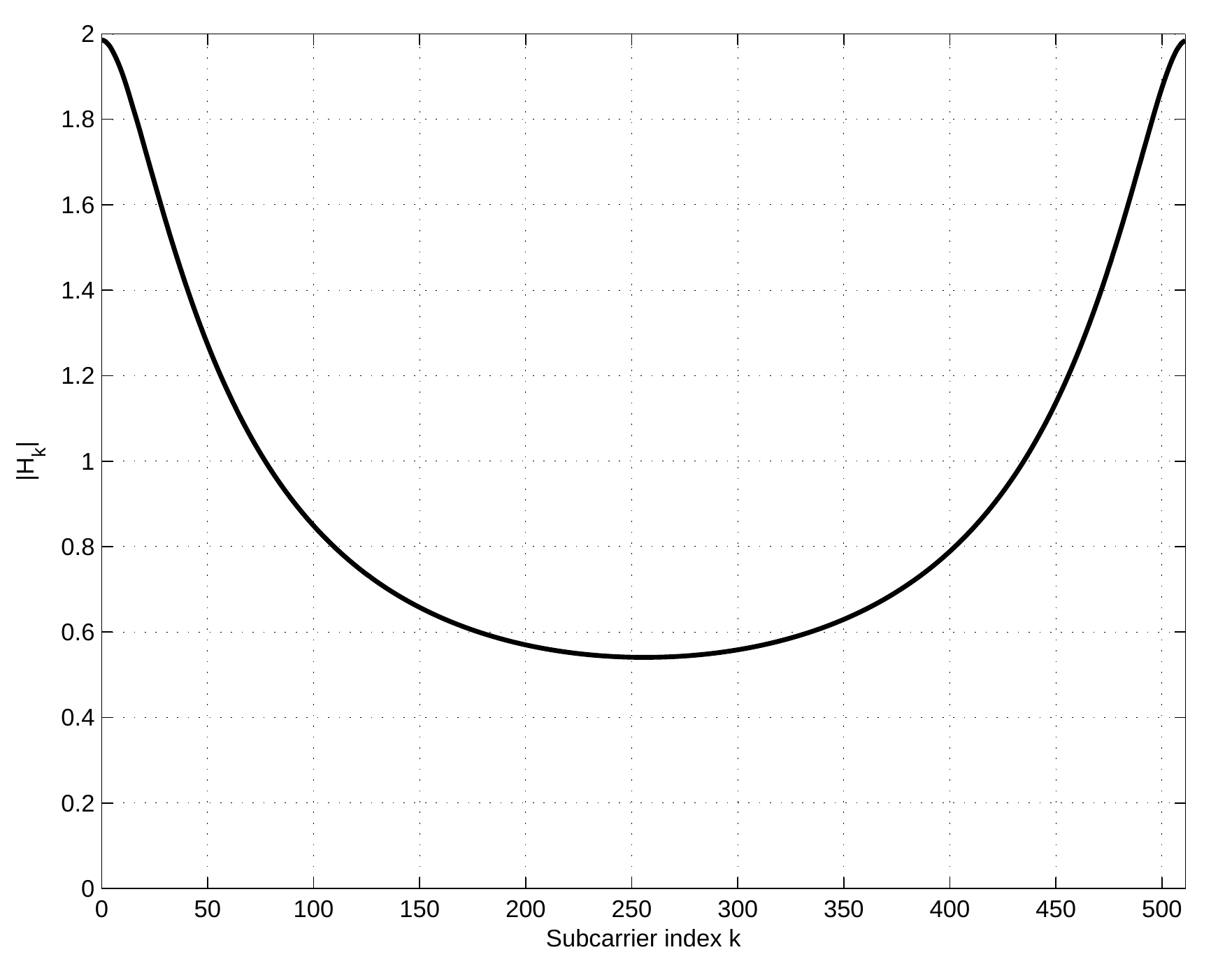}
\caption{Normalized frequency response for each subcarrier (rms delay spread D = 10 ns, 100 MHz sampling rate).}
\label{figchannel}
\end{center}
\end{figure}

As examples, we chose $\eta_{\rm OSNR}= 20$\,dB, $\eta_{\rm DSNR} =
32$\,dB, and AWGN channel. Figures 7 and 8 show the achievable data rate as a function of the
clipping ratio and the biasing ratio for DCO-OFDM and ACO-OFDM,
respectively. We see that for given $\eta_{\rm OSNR}$ and $\eta_{\rm
DSNR}$ values, a pair of optimum clipping ratio $\gamma^{\ddag}$ and
optimum biasing ratio $\varsigma^{\ddag}$ exist that maximize the
achievable data rate. It is worthwhile to point out that the optimum
biasing ratio $\varsigma^{\ddag}$ is different from
$\varsigma^{\star}$ (recall that $\varsigma^{\star}$ minimizes the
EVM).
If the system is only subject to the dynamic power constraint,
$\varsigma^{\ddag}$ should be equal to $\varsigma^{\star}$. If the
dominant constraint is the average power, $\varsigma^{\ddag}$ should
be less than or equal to $\varsigma^{\star}$ because reducing the
biasing ratio can make the signal average power lower.
We can obtain the optimum clipping ratio and biasing ratio for given
$\eta_{\rm OSNR}, \eta_{\rm DSNR}$ by
\begin{equation}
\label{eqoptclip} (\gamma^{\ddag}, \varsigma^{\ddag}) =
\underset{(\gamma,
\varsigma)}{\texttt{argmax}}\quad\mathcal{R}|_{\eta_{\rm OSNR},
\eta_{\rm DSNR}}
\end{equation}

\begin{figure}[!t]
\begin{center}
\includegraphics[width=5.0in]{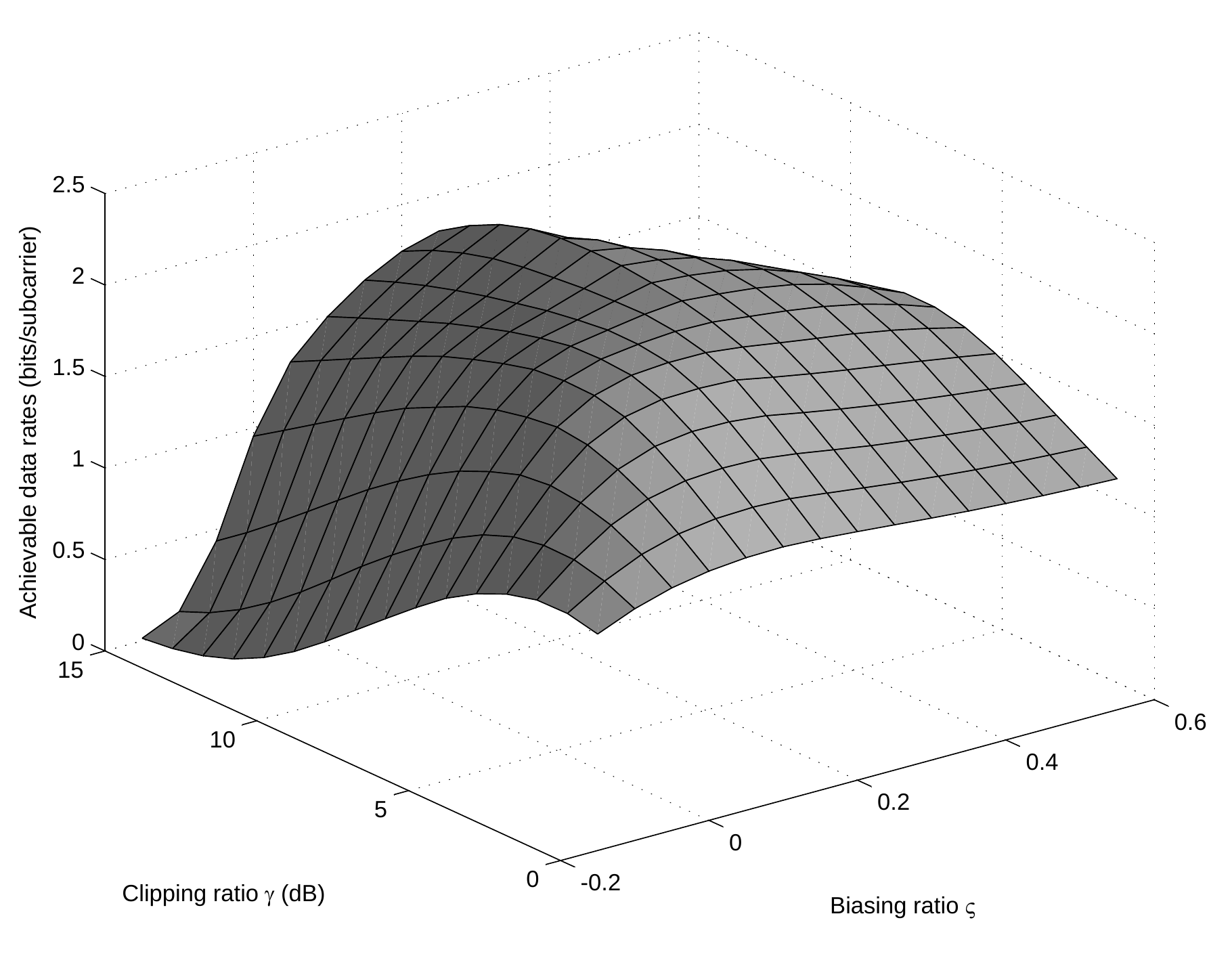}
\caption{Achievable data rate as a function of the clipping ratio and the biasing ratio for DCO-OFDM with $\eta_{OSNR}= 20 $ dB, $\eta_{DSNR} = 32$ dB, and AWGN channel.}
\label{figdco3d}
\end{center}
\end{figure}

\begin{figure}[!t]
\begin{center}
\includegraphics[width=5.0in]{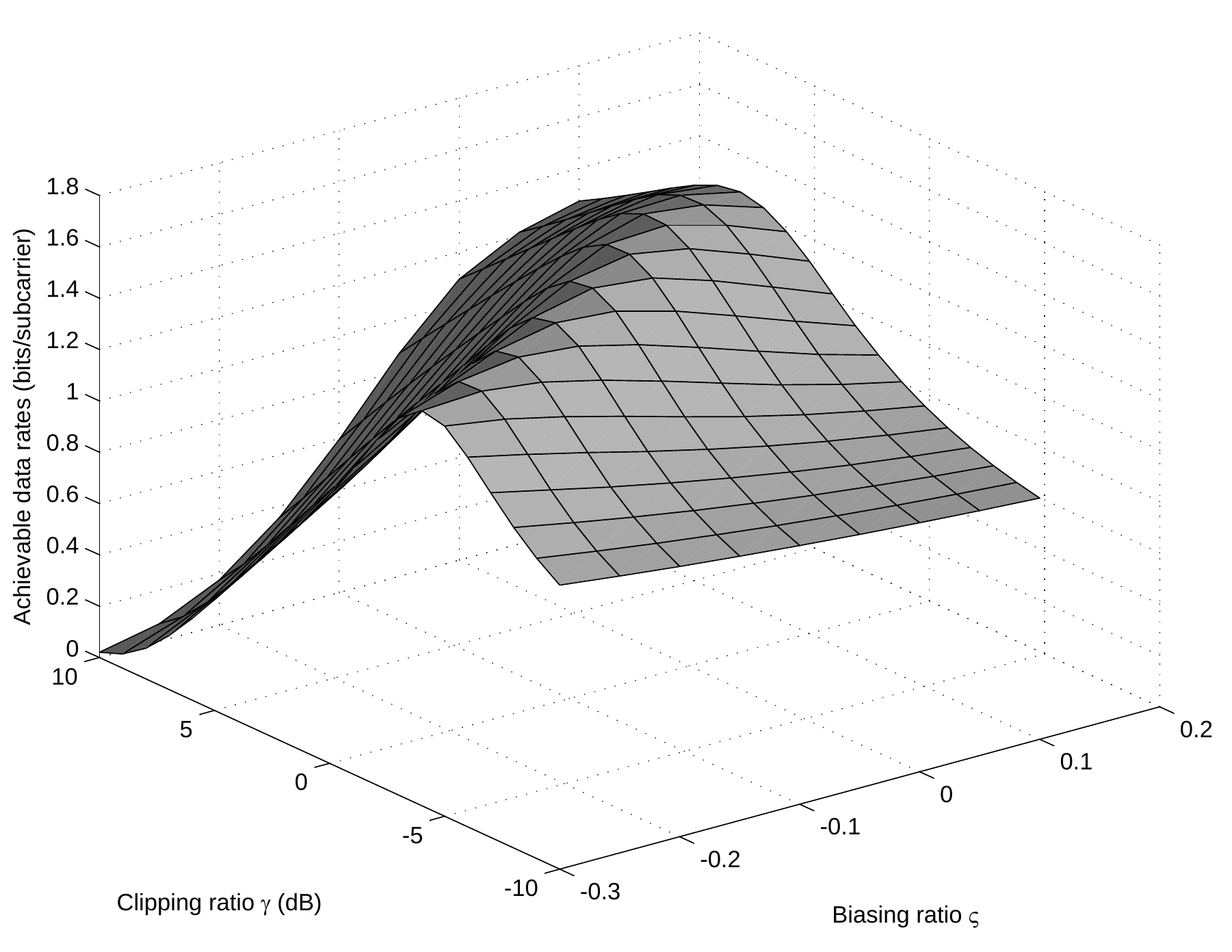}
\caption{Achievable data rate as a function of the clipping ratio and the biasing ratio for ACO-OFDM with $\eta_{OSNR}= 20$ dB, $\eta_{DSNR} = 32$ dB, and AWGN channel.}
\label{figaco3d}
\end{center}
\end{figure}

Figure 9a shows the optimal clipping ratio as a
function of $\eta_{\rm OSNR}$ for DCO-OFDM.
Figure 9b shows the optimum biasing ratio as a
function of $\eta_{\rm OSNR}$ for DCO-OFDM. Similar plots are shown
as Figure 10a,b for ACO-OFDM. In all cases,
$\eta_{\rm OSNR}$ varied from 0 to 25\,dB in step size of 1\,dB,
$\eta_{\rm DSNR}/\eta_{\rm OSNR}= 18$\,dB, and the channel was AWGN.
The main observation is, with a lower average optical power
constraint, the clipping ratio and the biasing ratio can be
increased to achieve higher data rates. Intuitively, when $\eta_{\rm
OSNR}$ is large, the channel noise has little effect and the
nonlinear distortion dominates.

\begin{figure}[!t]
\begin{center}
\includegraphics[width=5in]{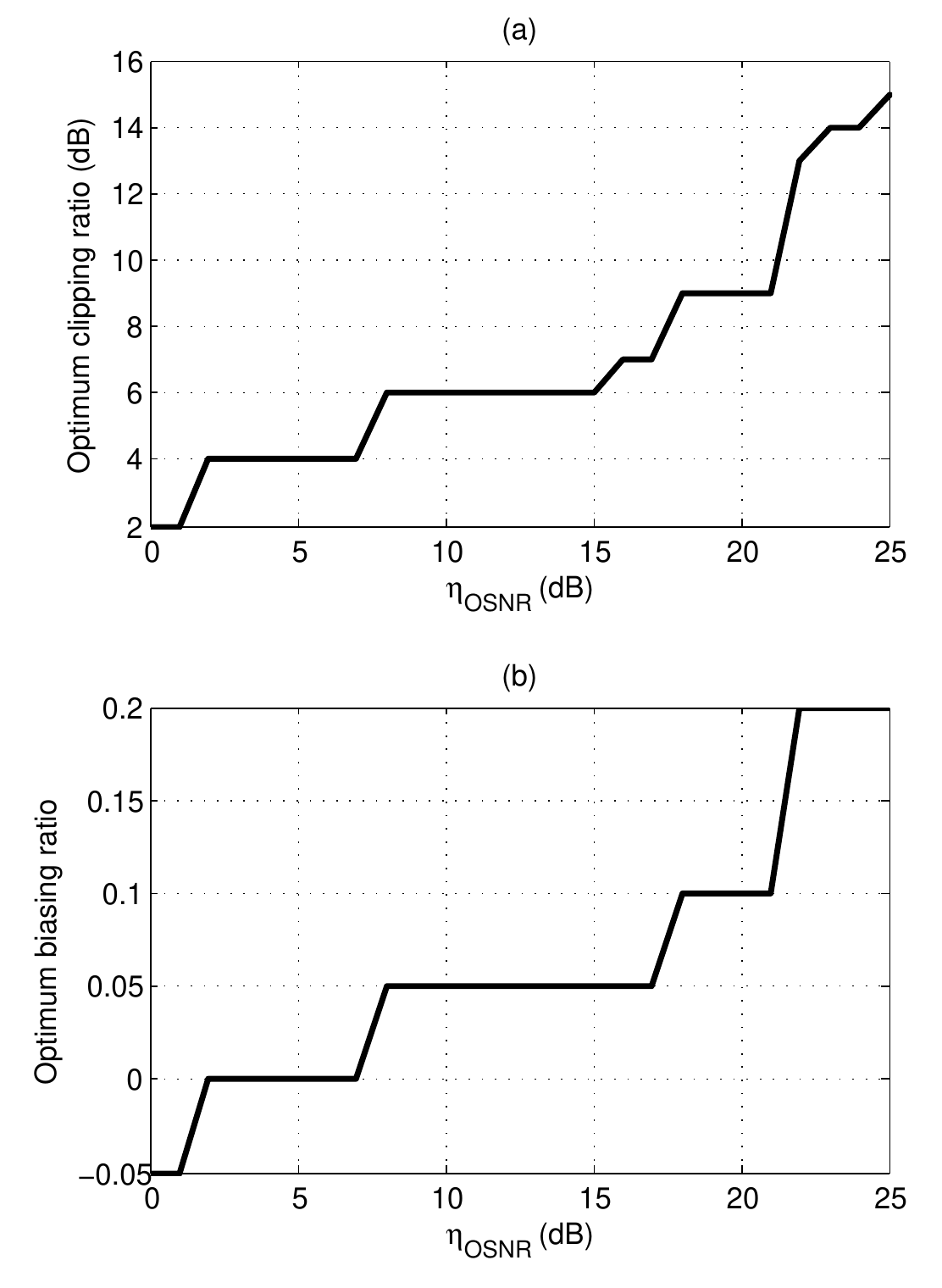}
\caption{Optimal clipping ratio and biasing ratio of DCO-OFDM for $\eta_{OSNR}= 0, 1,..., 25 $ dB (in step size of 1 dB), $\eta_{DSNR}/\eta_{OSNR}= 18 $ dB, and AWGN channel.}
\label{figdoubledco}
\end{center}
\end{figure}

\begin{figure}[!t]
\begin{center}
\includegraphics[width=5in]{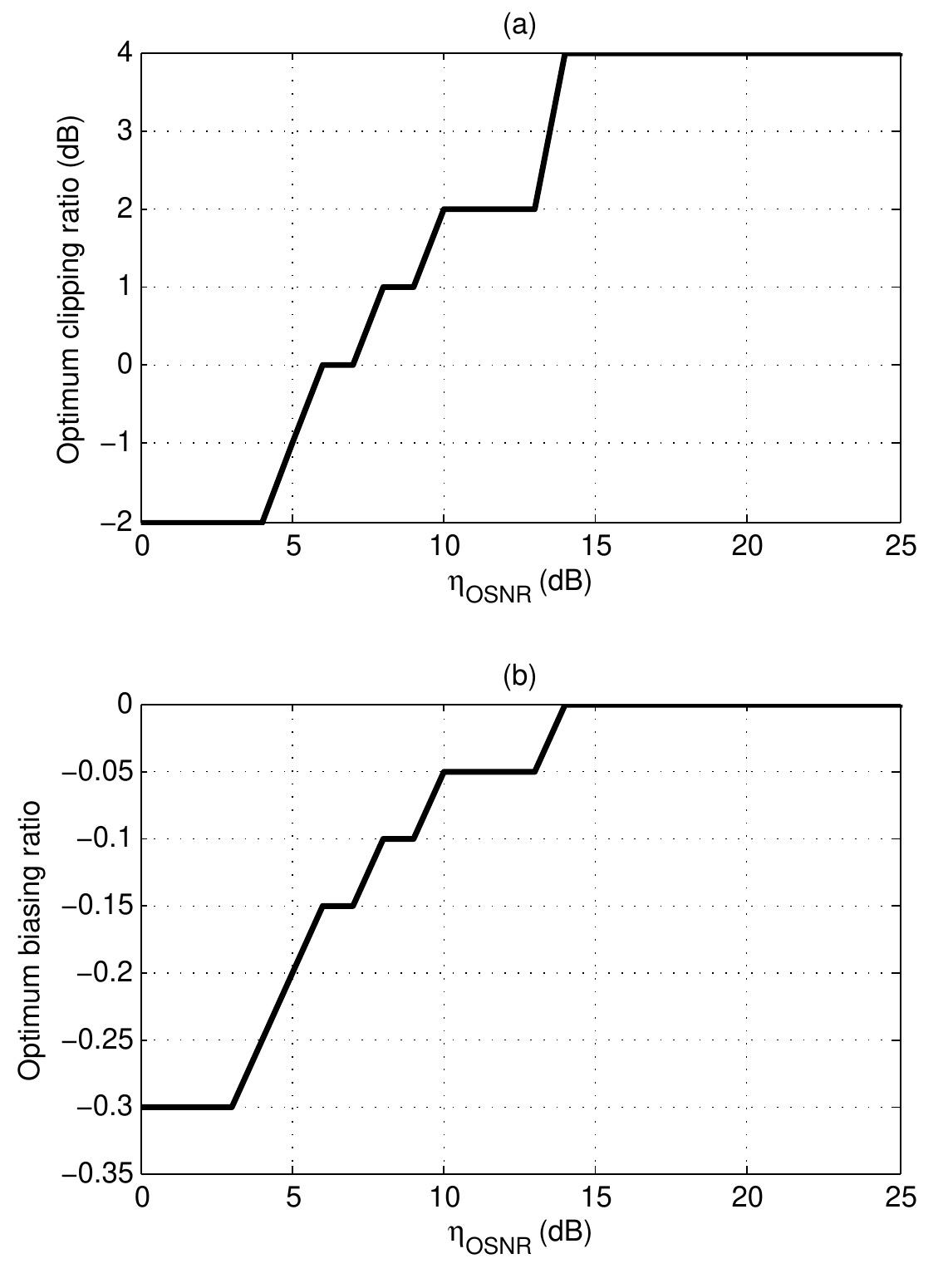}
\caption{Optimal clipping ratio and biasing ratio of ACO-OFDM for $\eta_{OSNR}= 0, 1,..., 25 $ dB (in step size of 1 dB), $\eta_{DSNR}/\eta_{OSNR}= 18 $ dB, and AWGN channel.}
\label{figdoubleaco}
\end{center}
\end{figure}
Next, we chose the ratio $\eta_{\rm DSNR}/\eta_{\rm OSNR}$ from
6\,dB, 12\,dB, and no $\eta_{\rm DSNR}$ constraints. For each pair
of $\eta_{\rm OSNR}, \eta_{\rm DSNR}$, AWGN channel, or
frequency-selective channel, we can calculate the optimum clipping
ratio $\gamma^{\ddag}$ and biasing ratio $\varsigma^{\ddag}$
according to Equation (\ref{eqoptclip}) and the corresponding
achievable data rates. Figures 11,
12, and 13 show the achievable data
rates with optimal clipping ratio and biasing ratio for the case
$\eta_{\rm DSNR}/\eta_{\rm OSNR}= 6$\,dB, $\eta_{\rm DSNR}/\eta_{\rm
OSNR}= 12$\,dB, and no $\eta_{\rm DSNR}$ constraint, respectively.
We observe that the performance of ACO-OFDM and DCO-OFDM depends on
the specific optical power constraints scenario. In general,
DCO-OFDM outperforms ACO-OFDM for all the cases. With the increase
of the ratio $\eta_{\rm DSNR}/\eta_{\rm OSNR}$, the average optical
power becomes the dominant constraint. The ACO-OFDM moves closer to
the DCO-OFDM.
\begin{figure}[!t]
\begin{center}
\includegraphics[width=5.0in]{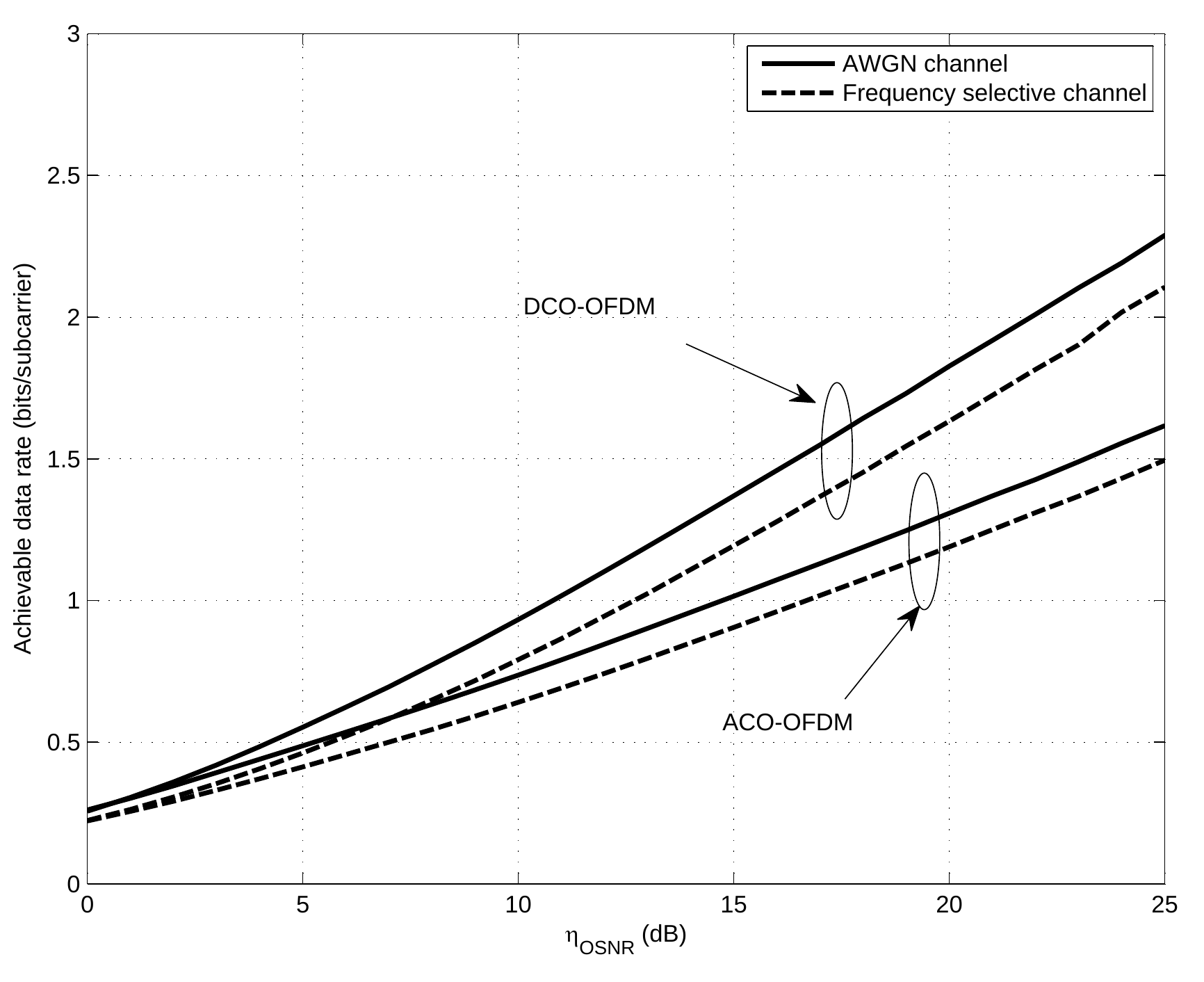}
\caption{Achievable data rate with optimal clipping ratio and optimal biasing ratio for $\eta_{OSNR}= 0, 1,..., 25 $ dB (in step size of 1 dB), and $\eta_{DSNR}/\eta_{OSNR}= 6 $ dB.}
\label{figoptrate6}
\end{center}
\end{figure}

\begin{figure}[htc]
\begin{center}
\includegraphics[width=5.0in]{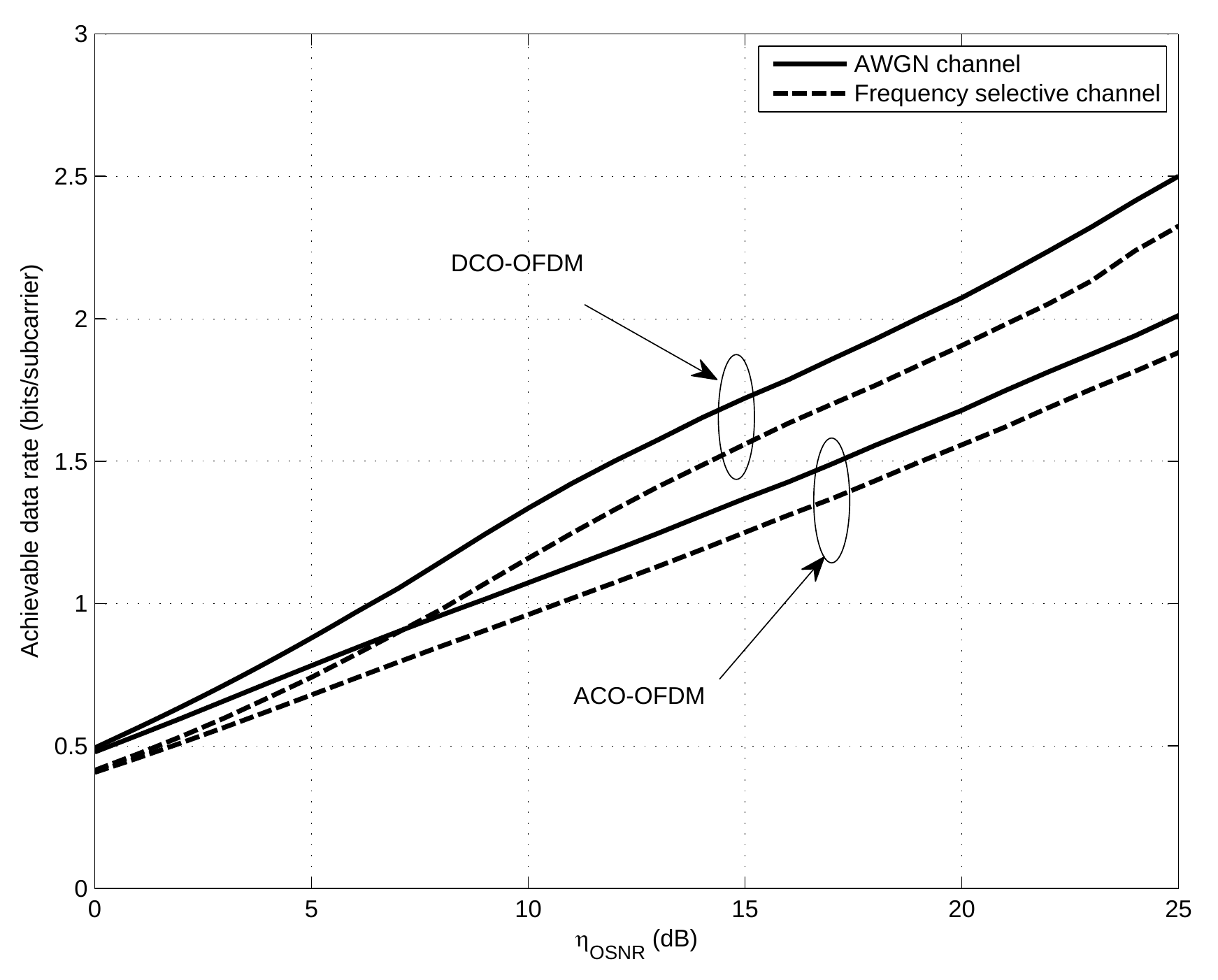}
\caption{Achievable data rate with optimal clipping ratio and biasing ratio for $\eta_{OSNR}= 0, 1,..., 25 $ dB (in step size of 1 dB), and $\eta_{DSNR}/\eta_{OSNR}= 12 $ dB.}
\label{figoptrate12}
\end{center}
\end{figure}

\begin{figure}[htc]
\begin{center}
\includegraphics[width=5.0in]{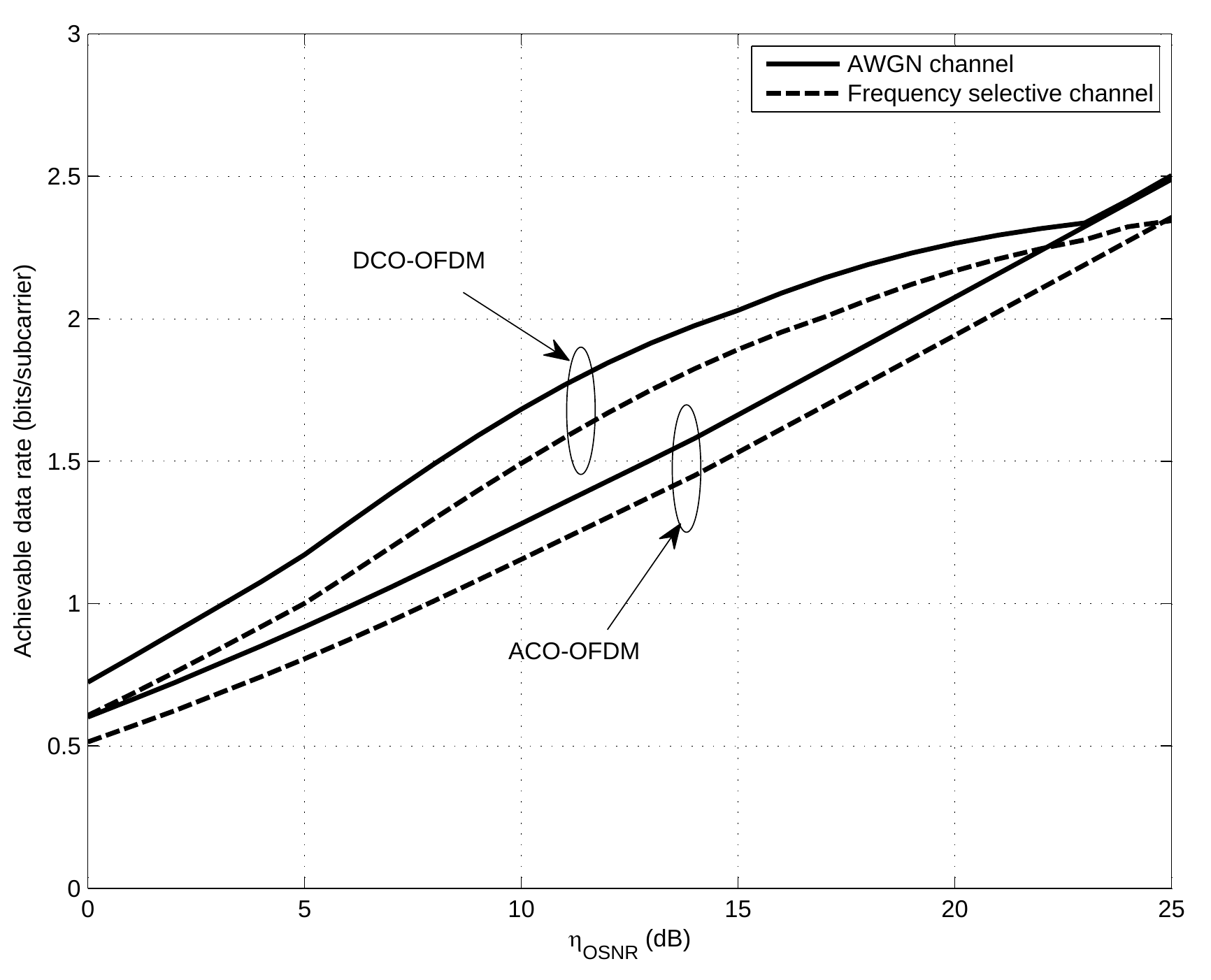}
\caption{Achievable data rate with optimal clipping ratio and biasing ratio for $\eta_{OSNR}= 0,1,..., 25 $ dB (in step size of 1 dB), and no $\eta_{DSNR}$ constraint.}
\label{figoptrate100}
\end{center}
\end{figure}

As seen in Figure 13, when there is no DSNR
constraint and the OSNR constraint is large, the DCO-OFDM curve
closely matches the ACO-OFDM curve.  This is in contrast to the
performance curves in Figures 11 and
12. The reason for the curve coincidence in
Figure 13 is twofold. First, we have already
discussed that the performance difference between DCO-OFDM and
ACO-OFDM is less when the OSNR constraint dominates, which is the
case for Figure 13. Second, the suddenness of the
convergence of the two curves can be explained by the fact that with
only OSNR constraint, there will be more flexibility in the signal
optimization to adjust the clipping ratio and biasing ratio to
achieve the best performance. That means that the achievable data
rates in the middle-OSNR region (5--22\,dB) are improved
significantly compared with Figures 11 and
12. However, for high-OSNR region (greater than
22\,dB), since the nonlinear distortion is negligible, the
improvement becomes less pronounced compared to
Figures 11 and 12. Therefore, the
transition from the middle-OSNR region to the high-OSNR region will
become sharper with only an OSNR constraint.

\section{Conclusions}
In this article, we analyzed the performance of the DCO-OFDM and
ACO-OFDM systems in terms of EVM, SDR, and achievable data rates
under both the average optical power and dynamic optical power
constraints. We numerically calculated the EVM and compared with the
corresponding lower bound. Both the theory and the simulation
results showed that ACO-OFDM can achieve the EVM lower bound. We
derived the achievable data rates for AWGN channel as well as
frequency-selective channel scenarios. We investigated the trade-off
between the optical power constraint and distortion. We analyzed the
optimum clipping ratio and biasing ratio and compared the
performance of two optical OFDM techniques. Numerical results showed
that DCO-OFDM outperforms the ACO-OFDM for all the optical power
constraint scenarios.

\section*{Competing interests}
The authors declare that they have no competing interests.

\section*{Acknowledgments}
This study was supported in part by the Texas Instruments DSP
Leadership University Program.

\section*{Appendix}
\subsection*{Proof that ${c^{\star}[n] = x^{(A)}[n]-
\bar{x}^{(A)}[n]}$ is optimum for Equation (\ref{eq_finalopt})}

Denote by $u(c[n])$ the objective function for the problem in
(\ref{eq_finalopt}):
\begin{equation}
u(c[n]) = \sum_{n=0}^{N/2-1} \left(c[n]\right)^2,
\end{equation}
and denote by $g_i(c[n])$ the $i$th constraint function for the
problem in (\ref{eq_finalopt}):
\begin{eqnarray}
g_i(c[n]) =\left\{\begin{array}{ccl}x^{(A)}[i] +
c[i]-2\gamma\sigma,\,\, &&{0 \leq i \leq \frac{N}{2}-1 }\\
-x^{(A)}[i-N/2] - c[i-N/2],\,\,&&  {\frac{N}{2} \leq i \leq
N-1}\end{array}\right..
\end{eqnarray}
Let $\mu_i$ denote the $i$th Kuhn--Tucker (KT) multiplier. We inter
that
\begin{eqnarray}
\nabla u(c[n]) = \left[2c[0], 2c[1], \dots, 2c[N/2-1]\right],
\end{eqnarray}
\begin{eqnarray}
\sum_{i=0}^{N-1}\mu_i\nabla g_i(c[n]) = \left[\mu_0 - \mu_{N/2},
\mu_i - \mu_{N/2+1}, \ldots, \mu_{N/2-1} - \mu_{N-1}\right].
\end{eqnarray}

Next, we prove that $c^{\star}[n] = x^{(A)}[n]- \bar{x}^{(A)}[n]$
satisfies the KT conditions \cite{kuhn1951nonlinear}.

Stationarity
\begin{equation}
\label{eq_ktsta} \nabla u(c[n]) + \sum_{i=0}^{N-1}\mu_i\nabla
g_i(c[n]) = 0.
\end{equation}

Primary feasibility
\begin{equation}
\label{eq_ktpf} \mu_i \geq 0, \quad \forall i = 0, 1, \ldots, N-1.
\end{equation}

Dual feasibility
\begin{equation}
\label{eq_ktdf} g_i(c[n]) \leq 0, \quad \forall i = 0, 1,\ldots,
N-1.
\end{equation}

Complementary slackness
\begin{equation}
\label{eq_ktcs} \mu_ig_i(c[n]) = 0, \quad \forall i = 0, 1, \ldots,
N-1.
\end{equation}

Substituting $c^{\star}[n]$ into Equation (\ref{eq_ktsta}), we
obtain,
\begin{equation}
\mu_i - \mu_{i+N/2} = -2x^{(A)}[i]+ 2\bar{x}^{(A)}[i], \quad i = 0,
1, \ldots, \frac{N}{2}-1.
\end{equation}
In order to satisfy all the other conditions
(\ref{eq_ktpf})--(\ref{eq_ktcs}), we can choose $\mu_i$ as follows

(1) if $x^{(A)}[i] = \bar{x}^{(A)}[i]$,
\begin{equation}
\mu_i = \mu_{i+N/2} = 0;
\end{equation}

(2) if $x^{(A)}[i] > 2\gamma\sigma$ and $\bar{x}^{(A)}[i] = 2\gamma\sigma$,
\begin{equation}
\mu_i = 2\bar{x}^{(A)}[i], \quad \mu_{i+N/2} = 2x^{(A)}[i];
\end{equation}

(3) if $x^{(A)}[i] < 0$ and $\bar{x}^{(A)}[i] = 0$,
\begin{equation}
\mu_i = -2x^{(A)}[i], \quad \mu_{i+N/2} = -2\bar{x}^{(A)}[i] .
\end{equation}
Therefore, there exits constants $\mu_i \, (i = 0, 1, \ldots, N-1)$
that make $c^{\star}[n] = x^{(A)}[n]- \bar{x}^{(A)}[n]$ satisfy the
KT conditions. It was shown in \cite{hanson1999invexity} that if the
objective function and the constraint functions are continuously
differentiable convex functions, KT conditions are sufficient for
optimality. It is obvious that $u$ and $g$ are all continuously
differentiable convex functions. Therefore, $c^{\star}[n]$ is
optimal for the minimization problem (\ref{eq_finalopt}).
%
%





%

\ifCLASSOPTIONcaptionsoff
  \newpage
\fi



%
\bibliographystyle{IEEEtran}
\bibliography{bmc_article}

\begin{thebibliography}{10}
\providecommand{\url}[1]{#1}
\csname url@samestyle\endcsname
\providecommand{\newblock}{\relax}
\providecommand{\bibinfo}[2]{#2}
\providecommand{\BIBentrySTDinterwordspacing}{\spaceskip=0pt\relax}
\providecommand{\BIBentryALTinterwordstretchfactor}{4}
\providecommand{\BIBentryALTinterwordspacing}{\spaceskip=\fontdimen2\font plus
\BIBentryALTinterwordstretchfactor\fontdimen3\font minus
  \fontdimen4\font\relax}
\providecommand{\BIBforeignlanguage}[2]{{%
\expandafter\ifx\csname l@#1\endcsname\relax
\typeout{** WARNING: IEEEtran.bst: No hyphenation pattern has been}%
\typeout{** loaded for the language `#1'. Using the pattern for}%
\typeout{** the default language instead.}%
\else
\language=\csname l@#1\endcsname
\fi
#2}}
\providecommand{\BIBdecl}{\relax}
\BIBdecl

\bibitem{zyu2012evm}
Z.~Yu, R.~J. Baxley, and G.~T. Zhou, ``Evm and achievable data rate analysis of
  clipped {OFDM} signals in visible light communication,'' \emph{EURASIP
  Journal on Wireless Communications and Networking}, vol. 2012, no. 321, 2012.

\bibitem{komine2003integrated}
T.~Komine and M.~Nakagawa, ``Integrated system of white led visible-light
  communication and power-line communication,'' \emph{IEEE Transactions on
  Consumer Electronics}, vol.~49, no.~1, pp. 71--79, 2003.

\bibitem{komine2004fundamental}
------, ``Fundamental analysis for visible-light communication system using led
  lights,'' \emph{IEEE Transactions on Consumer Electronics}, vol.~50, no.~1,
  pp. 100--107, 2004.

\bibitem{o2008visible}
D.~O'Brien, L.~Zeng, H.~Le-Minh, G.~Faulkner, J.~W. Walewski, and S.~Randel,
  ``Visible light communications: Challenges and possibilities,'' in \emph{IEEE
  19th International Symposium on Personal, Indoor and Mobile Radio
  Communications, 2008. PIMRC 2008}.\hskip 1em plus 0.5em minus 0.4em\relax
  IEEE, 2008, pp. 1--5.

\bibitem{elgala2011indoor}
H.~Elgala, R.~Mesleh, and H.~Haas, ``Indoor optical wireless communication:
  potential and state-of-the-art,'' \emph{IEEE Communications Magazine},
  vol.~49, no.~9, pp. 56--62, 2011.

\bibitem{elgala2007ofdm}
H.~Elgala, R.~Mesleh, H.~Haas, and B.~Pricope, ``Ofdm visible light wireless
  communication based on white leds,'' in \emph{IEEE 65th Vehicular Technology
  Conference, 2007. VTC2007-Spring.}\hskip 1em plus 0.5em minus 0.4em\relax
  Ieee, 2007, pp. 2185--2189.

\bibitem{hranilovic2005design}
S.~Hranilovic, ``On the design of bandwidth efficient signalling for indoor
  wireless optical channels,'' \emph{International Journal of Communication
  Systems}, vol.~18, no.~3, pp. 205--228, 2005.

\bibitem{gonzalez2006adaptive}
O.~Gonzalez, R.~Perez-Jimenez, S.~Rodriguez, J.~Rabadan, and A.~Ayala,
  ``Adaptive ofdm system for communications over the indoor wireless optical
  channel,'' \emph{IEE Proceedings-Optoelectronics}, vol. 153, p. 139, 2006.

\bibitem{armstrong2006power}
J.~Armstrong and A.~J. Lowery, ``Power efficient optical {OFDM},''
  \emph{Electronics Letters}, vol.~42, no.~6, pp. 370--372, 2006.

\bibitem{armstrong2009ofdm}
J.~Armstrong, ``Ofdm for optical communications,'' \emph{Journal of Lightwave
  Technology}, vol.~27, no.~3, pp. 189--204, 2009.

\bibitem{fernando2011flip}
N.~Fernando, Y.~Hong, and E.~Viterbo, ``Flip-ofdm for optical wireless
  communications,'' in \emph{IEEE Information Theory Workshop, 2011}.\hskip 1em
  plus 0.5em minus 0.4em\relax IEEE, 2011, pp. 5--9.

\bibitem{carruthers1996multiple}
J.~Carruthers and J.~Kahn, ``Multiple-subcarrier modulation for nondirected
  wireless infrared communication,'' \emph{IEEE Journal on Selected Areas in
  Communications}, vol.~14, no.~3, pp. 538--546, 1996.

\bibitem{tellado2000multicarrier}
J.~Tellado, \emph{Multicarrier modulation with low PAR: applications to DSL and
  wireless}.\hskip 1em plus 0.5em minus 0.4em\relax Springer, 2000, vol. 587.

\bibitem{elgala2009non}
H.~Elgala, R.~Mesleh, and H.~Haas, ``Non-linearity effects and predistortion in
  optical {OFDM} wireless transmission using {LED}s,'' \emph{International
  Journal of Ultra Wideband Communications and Systems}, vol.~1, no.~2, pp.
  143--150, 2009.

\bibitem{banelli2000theoretical}
P.~Banelli and S.~Cacopardi, ``Theoretical analysis and performance of ofdm
  signals in nonlinear awgn channels,'' \emph{IEEE Transactions on
  Communications}, vol.~48, no.~3, pp. 430--441, 2000.

\bibitem{ochiai2002performance}
H.~Ochiai and H.~Imai, ``Performance analysis of deliberately clipped ofdm
  signals,'' \emph{IEEE Transactions on Communications}, vol.~50, no.~1, pp.
  89--101, 2002.

\bibitem{ochiai2003performance}
H.~Ochiai, ``Performance analysis of peak power and band-limited ofdm system
  with linear scaling,'' \emph{IEEE Transactions on Wireless Communications},
  vol.~2, no.~5, pp. 1055--1065, 2003.

\bibitem{peng2006capacity}
F.~Peng and W.~Ryan, ``On the capacity of clipped ofdm channels,'' in
  \emph{Information Theory, 2006 IEEE International Symposium on}.\hskip 1em
  plus 0.5em minus 0.4em\relax IEEE, 2006, pp. 1866--1870.

\bibitem{armstrong2008comparison}
J.~Armstrong and B.~Schmidt, ``Comparison of asymmetrically clipped optical
  ofdm and dc-biased optical ofdm in awgn,'' \emph{IEEE Communications
  Letters}, vol.~12, no.~5, pp. 343--345, 2008.

\bibitem{elgala2009study}
H.~Elgala, R.~Mesleh, and H.~Haas, ``A study of {LED} nonlinearity effects on
  optical wireless transmission using {OFDM},'' in \emph{International
  Conference on Wireless and Optical Communications Networks}.\hskip 1em plus
  0.5em minus 0.4em\relax IEEE, 2009, pp. 1--5.

\bibitem{mesleh2011performance}
R.~Mesleh, H.~Elgala, and H.~Haas, ``On the performance of different ofdm based
  optical wireless communication systems,'' \emph{Journal of Optical
  Communications and Networking}, vol.~3, no.~8, pp. 620--628, 2011.

\bibitem{li2007channel}
X.~Li, R.~Mardling, and J.~Armstrong, ``Channel capacity of im/dd optical
  communication systems and of aco-ofdm,'' in \emph{IEEE International
  Conference on Communications, 2007. ICC'07.}\hskip 1em plus 0.5em minus
  0.4em\relax IEEE, 2007, pp. 2128--2133.

\bibitem{wilson2009transmitter}
S.~Wilson and J.~Armstrong, ``Transmitter and receiver methods for improving
  asymmetrically-clipped optical {OFDM},'' \emph{IEEE Transactions on Wireless
  Communications}, vol.~8, no.~9, pp. 4561--4567, 2009.

\bibitem{bussgang1952crosscorrelation}
J.~Bussgang, ``Crosscorrelation functions of amplitude-distorted gaussian
  signals,'' \emph{NeuroReport}, vol.~17, no.~2, 1952.

\bibitem{davenport1958introduction}
W.~Davenport and W.~Root, \emph{An introduction to the theory of random signals
  and noise}.\hskip 1em plus 0.5em minus 0.4em\relax IEEE Press, 1987.

\bibitem{abramowitz1964handbook}
M.~Abramowitz and I.~Stegun, \emph{Handbook of mathematical functions with
  formulas, graphs, and mathematical tables}.\hskip 1em plus 0.5em minus
  0.4em\relax Dover publications, 1964, vol.~55, no. 1972.

\bibitem{kahn1997wireless}
J.~Kahn and J.~Barry, ``Wireless infrared communications,'' \emph{Proceedings
  of the IEEE}, vol.~85, no.~2, pp. 265--298, 1997.

\bibitem{CVX}
M.~Grant and S.~Boyd, ``{CVX}: Matlab software for disciplined convex
  programming, version 1.21,'' \url{http://cvxr.com}, Apr. 2011.

\bibitem{kuhn1951nonlinear}
H.~Kuhn and A.~Tucker, ``Nonlinear programming,'' in \emph{Proceedings of the
  second Berkeley symposium on mathematical statistics and probability},
  vol.~1.\hskip 1em plus 0.5em minus 0.4em\relax California, 1951, pp.
  481--492.

\bibitem{hanson1999invexity}
M.~Hanson, ``Invexity and the kuhn--tucker theorem,'' \emph{Journal of
  mathematical analysis and applications}, vol. 236, no.~2, pp. 594--604, 1999.

\end{thebibliography}


%

%
%
%




\end{document}